\documentclass[conference]{IEEEtran}
\IEEEoverridecommandlockouts
\usepackage{amsmath,amssymb,amsfonts}
\usepackage{algorithmic}
\usepackage{graphicx}
\graphicspath{{figures/},{pics}}
\usepackage{subfigure}
\usepackage{setspace}
\usepackage{textcomp}
\usepackage{xcolor}
\def\BibTeX{{\rm B\kern-.05em{\sc i\kern-.025em b}\kern-.08em
    T\kern-.1667em\lower.7ex\hbox{E}\kern-.125emX}}
\begin{document}

\title{Image Reconstruction for Multi-frequency Electromagnetic Tomography  based on Multiple Measurement Vector Model\\
\thanks{This work of Jinxi Xiang was supported in part by the China Scholarship Council under Grant 201906210259.}
}

\author{\IEEEauthorblockN{Jinxi Xiang}
\IEEEauthorblockA{\textit{University of Edinburgh}\\
Edinburgh, U.K. \\
\textit{Tsinghua University} \\
Beijing, China \\
xiangjx16@mails.tsinghua.edu.cn}
\and
\IEEEauthorblockN{Zhou Chen}
\IEEEauthorblockA{\textit{School of Engineering} \\
	\textit{University of Edinburgh}\\
	Edinburgh, U.K. \\
	zhou.chen@ed.ac.uk}
\and
\IEEEauthorblockN{Yonggui Dong}
\IEEEauthorblockA{\textit{Department of Precision Instrument} \\
\textit{Tsinghua University}\\
Beijing, China \\
dongyg@mail.tsinghua.edu.cn}
\and
\IEEEauthorblockN{Yunjie Yang}
\IEEEauthorblockA{\textit{School of Engineering} \\
	\textit{University of Edinburgh}\\
	Edinburgh, U.K. \\
	y.yang@ed.ac.uk}}

\maketitle

\begin{abstract}
Imaging the bio-impedance distribution of a biological sample can provide understandings about the sample's electrical properties which is an important indicator of physiological status. This paper presents a multi-frequency electromagnetic tomography (mfEMT) technique for biomedical imaging. The system consists of 8 channels of gradiometer coils with adjustable sensitivity and excitation frequency. To exploit the frequency correlation among each measurement, we reconstruct multiple frequency data simultaneously based on the Multiple Measurement Vector (MMV) model. The MMV problem is solved by using a sparse Bayesian learning method that is especially effective for sparse distribution. Both simulations and experiments have been conducted to verify the performance of the method. Results show that by taking advantage of multiple measurements, the proposed method is more robust to noisy data for ill-posed problems compared to the commonly used single measurement vector model.
\end{abstract}


\section{Introduction}

Biological tissues show complex impedance over a range of frequencies excited by time-varying electromagnetic fields \cite{o2015non}. Among the frequency range, $\beta$ dispersion operates between hundreds to megahertz, showing prominent frequency-dependent electrical properties \cite{pliquett2010bioimpedance}. This principle has been extensively explored in food agricultural produce \cite{soltani2011evaluating},  cell culture \cite{justice2011process} and clinic diagnosis \cite{xiao2018multi}.  

The most straightforward way of bio-impedance measurement is to contact or penetrate the sample by electrodes and then measure the transimpedance between the electrodes. For example, Yang and Jia reported a multi-frequency electrical impedance tomography (mfEIT) for biomedical imaging in \cite{yang2017multi}. However, in some situations, contact electrodes may cause problems due to contact impedance that could be considerable and variable with surface condition. Moreover, it maybe difficult or even impossible to place electrodes on certain test samples intrusively due to contamination or erosion issues. 

Multi-frequency Electromagnetic Tomography (mfEMT) is a contactless and non-invasive imaging technique. It measures bio-impedance by using multi-frequency excitations with components spreading across the bandwidth of interest  \cite{o2015non}.  Its potential application in intracranial hemorrhage detection was investigated in \cite{xiao2018multi}. A differential electromagnetic probe was designed for cervical tissue measurements in \cite{wang2017magnetic}. mfEMT is, in general, more challenging than electrical impedance tomography in terms of measurement because the higher sensitivity required for weak signal response.

There are two imaging modes of EMT, i.e. Time-Difference (TD) imaging and Frequency-Difference (FD) imaging \cite{jiang2018capacitively}. FD imaging utilizes the measurement between two frequencies and overcomes the limitation of the TD imaging that requires a reference data set. Commonly,  multi-frequency FD image reconstruction is to treat measurements independently and then solve them individually. This is also known as Single Measurement Vector (SMV) \cite{yang2011multiple}. In this way,  the correlations between measurements at different frequencies are overlooked.

In the field of compressive sensing, it has been shown that compared to the SMV model, the successful recovery rate of signal can be greatly improved by using Multipl Measurement Vectors (MMV) \cite{ziniel2012efficient}. Existing MMV applications include the direction of arrival estimation \cite{yang2012off} and magnetic resonance imaging \cite{majumdar2011accelerating}. Motivated by this, in this work, we apply the MMV model in mfEMT image reconstruction. The fundamental idea is that the spatial conductivity distribution of objects is highly correlated at different excitation frequencies. The MMV model exploits this kind of correlation and makes use of it to obtain a better-reconstructed image. Mathematically, this leads to a constrained optimization problem where the objective function promotes the signal's group-sparsity as well as its rank-deficiency. In the MMV model, the signal and noise statistics are learned directly from the data by sparse Bayesian learning that incorporates the frequency and spatial correlation of conductivity distribution into a prior. We then demonstrate the improvement of the proposed method through both simulations and experiments.

\section{Multi-frequency Electromagnetic Tomography System}
\label{sec:EMT}
\subsection{Physical Principle}

\begin{figure}[tb]
	\centering
	\includegraphics[width = 3in]{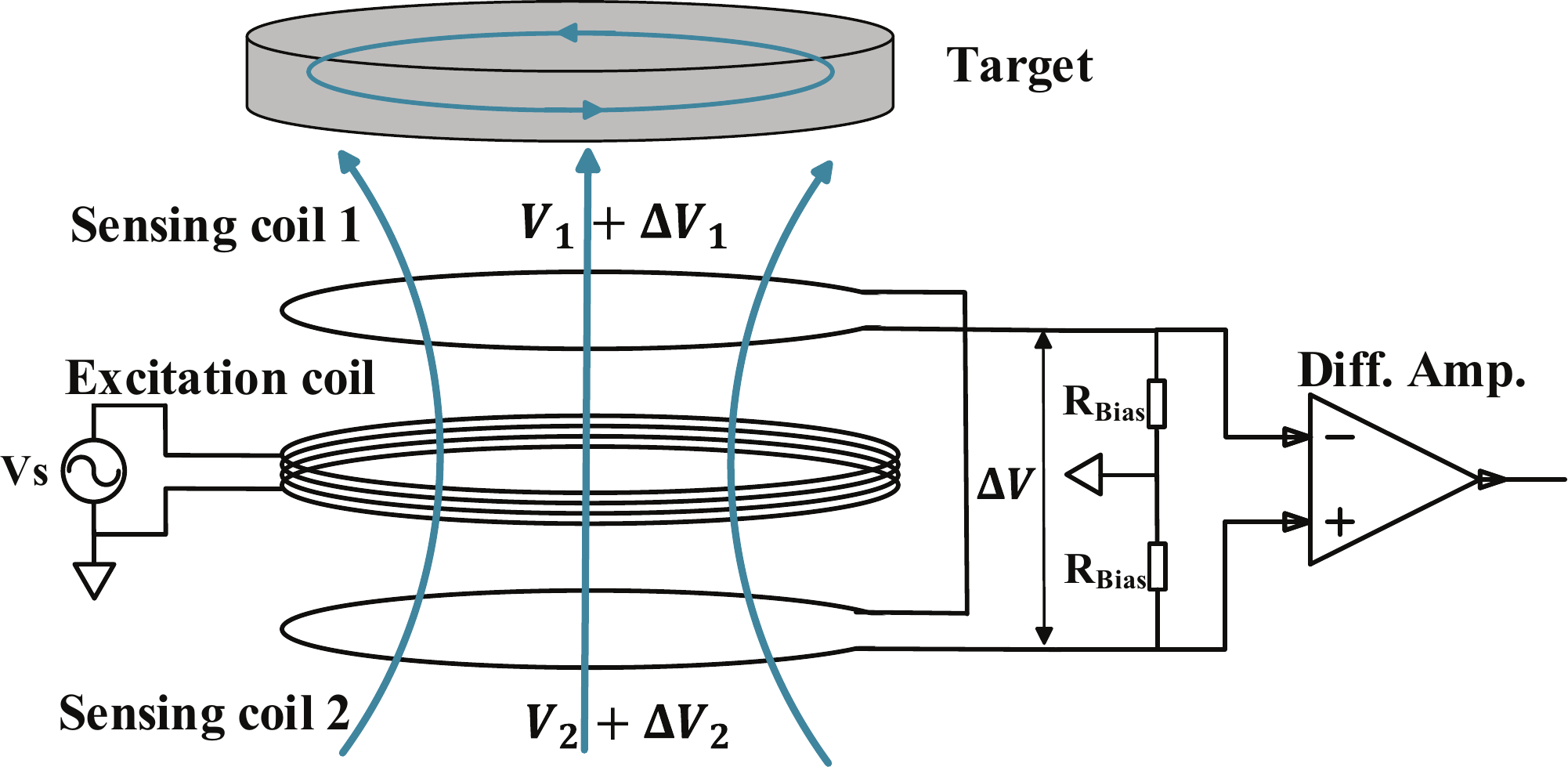}
	\caption{The configuration the gradiometer coil.}
	\label{fig:GradCoil}
\end{figure}

In mfEMT, a sample is placed within an alternating excitation magnetic field (primary field) that is produced by a current flowing in a coil. In return, an electrical field in the sample is generated which will cause eddy current and thus, a secondary magnetic filed can be measured externally. Due to the small conductivity of biological samples, the secondary field is weak. Therefore, it is suggested that the primary field should be eliminated  before measuring the secondary field. Previously, we reported that the gradiometer coil (see Fig.\ref{fig:GradCoil}) with differential structure improves the sensitivity greatly \cite{xiang2019design}. 

Based on the quasi-static approximation of eddy current, the sensitivity of gradiometer coil is governed by \cite{xiang2019design}:

\begin{equation}
	s_{g}=\frac{\Delta \varphi_{g}}{\Delta \sigma}=-\frac{V_{0}\left(P_{1}-P_{2}\right) \omega \mu_{0}}{V_{\mathrm{res}}}
	\label{eq:sensitivity}
\end{equation} 
where $\Delta \varphi_{g}$ denotes the phase response of secondary field caused by conductivity change $\Delta \sigma$. $V_0$ is the voltage produced by the primary magnetic field.  $P_1$ and $P_2$ are the geometrical factors that relate to the size and shape of the target and its position relative to the coil. $\omega$ represents the angular frequency of excitation signal and $\mu_0$ is the permeability in free space. $V_{\mathrm{res}}$, the key tunning parameter of sensitivity, is the residual voltage caused by the imbalance of two receiver coils.

\subsection{System Structure}

\begin{figure}[tbp]
	\centering
	\includegraphics[width = 3.2in]{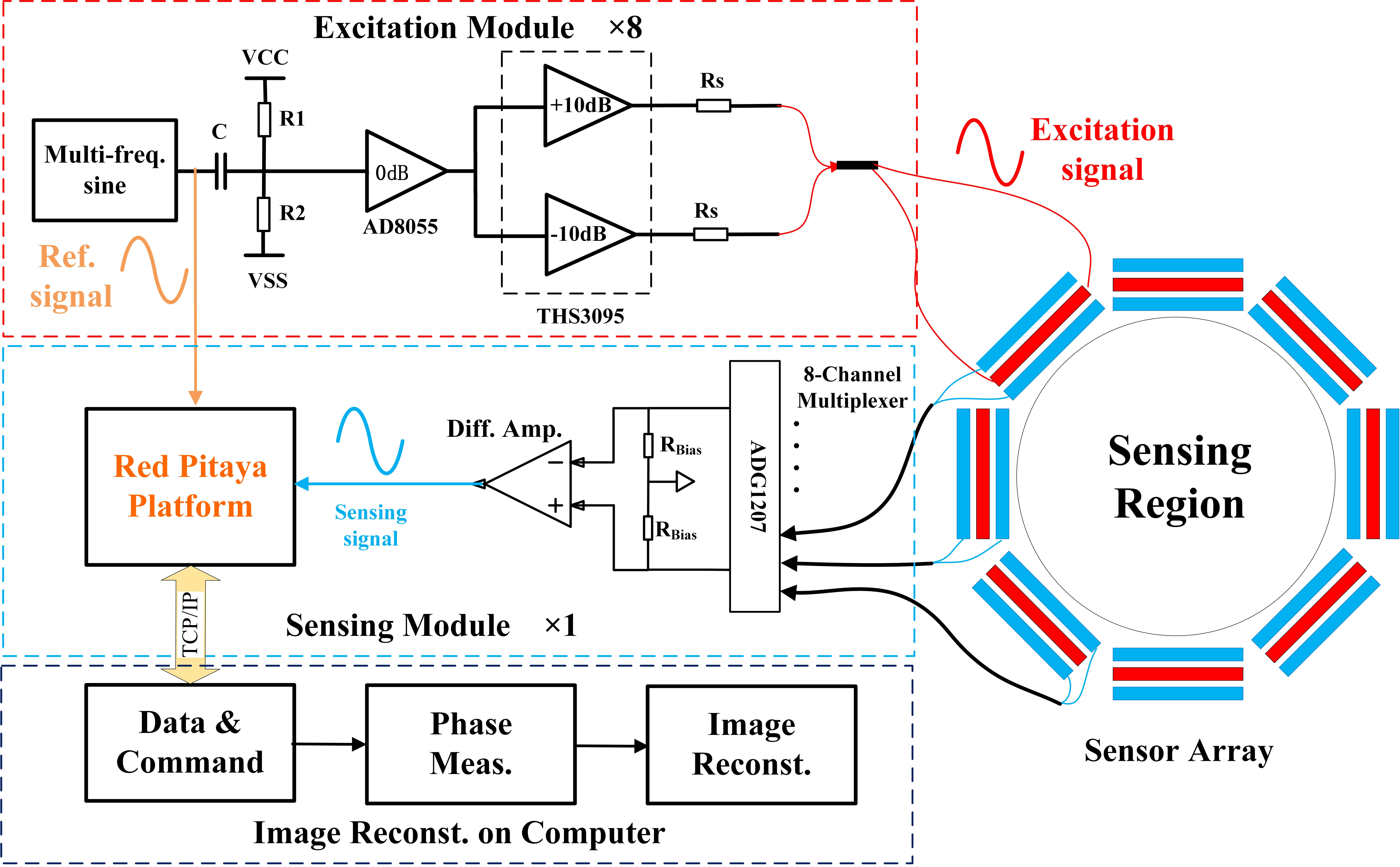}
	\caption{The 8-channel mfEMT system using gradiometer coils.}
	\label{fig:SystemRedPitaya}
\end{figure}

The mfEMT system (see Fig. \ref{fig:SystemRedPitaya}) comprises four modules :
\begin{itemize}
	\item[(1)]  sensor array consisting of 8 gradiometer coils;
	\item[(2)]  excitation modules to drive gradiometer coils;
	\item[(3)]  signal sensing modules and data acquisition based on Red Pitaya platform \cite{REDPITAYA};
	\item[(4)] 	phase measurement by Fast Fourier Transform (FFT) and image reconstruction in computer.
\end{itemize}

The operation principle is as follows. First of all, one of the 8 excitation channels is  enabled and the rest of the other 7 excitation channels are disabled as open circuits. Secondly, multi-frequency sine waves are generated by  Red Pitaya, an open-source hardware platform with dual fast (125 MSPS) ADCs and DACs.  Then,  8 differential sensing coils are sequentially selected and the sensing signals are multiplexed to the sensing module to measure the signal phase difference with the excitation signal.  One completed scan consists of 64 (8$\times$8) phase values covering all the excitation/sensing coil combinations. An image of the conductivity distribution is then reconstructed with the proposed MMV model solved by sparse Bayesian learning (see Section \ref{sec:MMV}).

The linear model describing phase change $\Delta \boldsymbol{\varphi}$  and the conductivity change $\Delta \boldsymbol{\sigma}$ is generally expressed as:
\begin{equation}
\Delta \boldsymbol{\varphi} = \mathbf{J} \Delta \boldsymbol{\sigma}
\label{eq:smatrix}
\end{equation}

Sensitivity matrix $\mathbf{J}$ is maps the conductivity distribution to phase values.  It reveals by previous work \cite{rosell2001sensitivity} that the sensitivity  at any pixel in the sensing region is the dot product of the two magnetic fields, which is expressed as: 
\begin{equation}
	\mathbf{J}(\Omega) = k ( \mathbf{B_1} \cdot \mathbf{B_2} ) 
	\label{eq:magnetic_field_dot}
\end{equation}
where  $\mathbf{B_1}$ is the magnetic field produced by a current $I_1$ injected into the excitation coil; $\mathbf{B_2}$ is the magnetic field produced by a current $I_2$ injected into the differential sensing coil; $k$ is a coefficient. The discretized mesh of $\mathbf{J}$ is $64\times64$. 

\begin{figure}[btp]
	\centering
	\includegraphics[width = 2.8in]{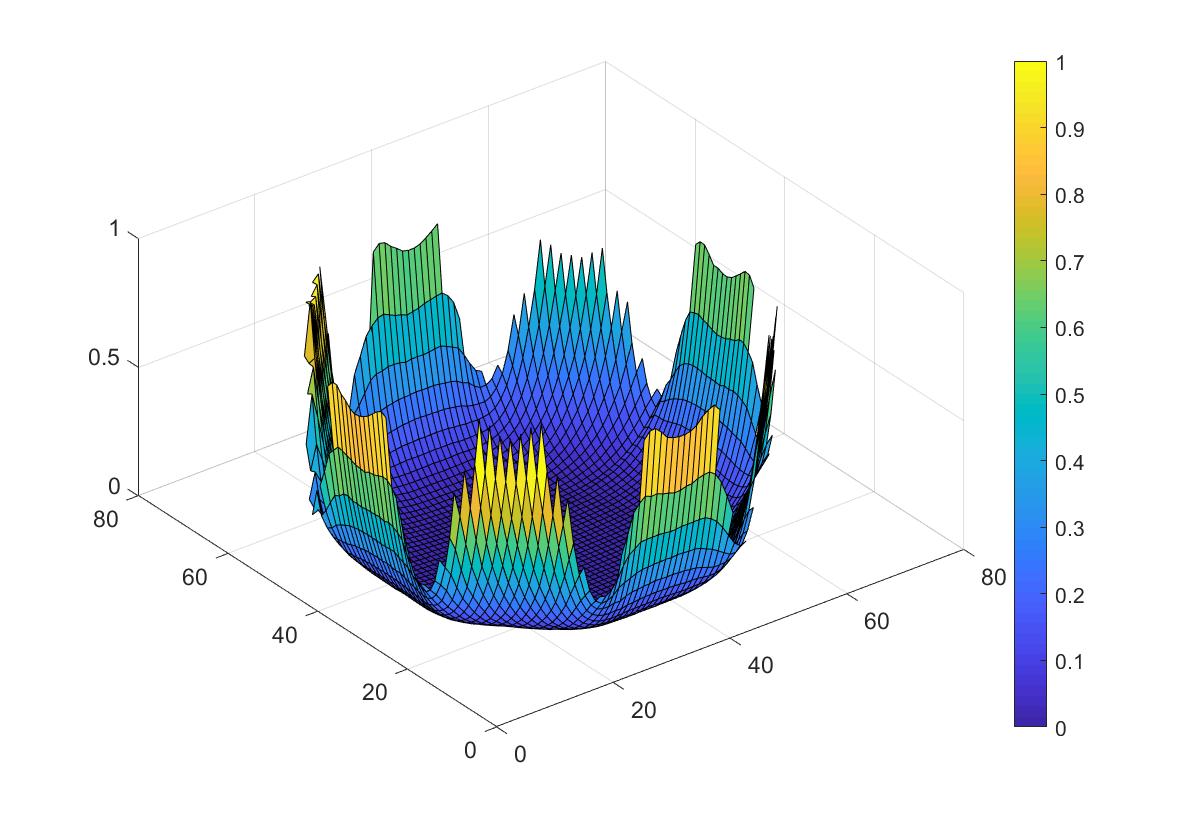}
	\caption{The sensitivity map $\mathbf{J}$ of the  8-channel mfEMT.}
	\label{fig:sMap}
\end{figure}

The normalized sensitivity map (summation of rows) is shown in Fig. \ref{fig:sMap}. It is observed that the largest sensitivity locates near the coils and smallest sensitivity appears at the central region. Other electrical tomographies such as electrical impedance tomography (EIT) and electrical capacitance tomography (ECT) share similar patterns.

\section{Multiple Measurement Vectors Framework using Sparse Bayesian Learning}
\label{sec:MMV}

To solve the mfEMT image reconstruction problem, the commonly used SMV framework considers the multiple equations independently and separately, as shown in Eq. (\ref{eq:SMV})

\begin{equation}
	\left\{
	\begin{array}{lr}
		\mathbf{y}_1 & =\mathbf{J} \boldsymbol{\sigma}_1+\mathbf{v}_1 \\
		\mathbf{y}_2 & =\mathbf{J} \boldsymbol{\sigma}_2+\mathbf{v}_2 \\
		\vdots &  \\
		\mathbf{y}_L & =\mathbf{J} \boldsymbol{\sigma}_L+\mathbf{v}_L
	\end{array}
	\right.
	\label{eq:SMV}
\end{equation}
where $\{\mathbf{y}_1, \mathbf{y}_2, \dots, \mathbf{y}_L\}$, $\mathbf{y}_i \in \mathbb{R}^{N\times 1}$ ($i = 1,2,\dots L$) denotes a sequence of measurements at different frequencies;  $\mathbf{J} \in \mathbb{R}^{N \times M}(N \ll M)$ represents the sensitivity matrix;   $\{\boldsymbol{\sigma}_1, \boldsymbol{\sigma}_2, \dots, \boldsymbol{\sigma}_L\}$, $\boldsymbol{\sigma}_i \in \mathbb{R}^{M\times 1}$ ($i = 1,2,\dots L$)  are the conductivity vectors to be solved and $\{\mathbf{v}_1, \mathbf{v}_2, \dots, \mathbf{v}_L\}$, $\mathbf{v}_i \in \mathbb{R}^{N\times 1}$ ($i = 1,2,\dots L$) are unknown noise vectors.
 
\begin{figure}[btp]
	\centering
	\subfigure[Solution $\boldsymbol{\sigma}$ in Eq.(\ref{eq:MMV1})]{\includegraphics[width = 1.9 in]{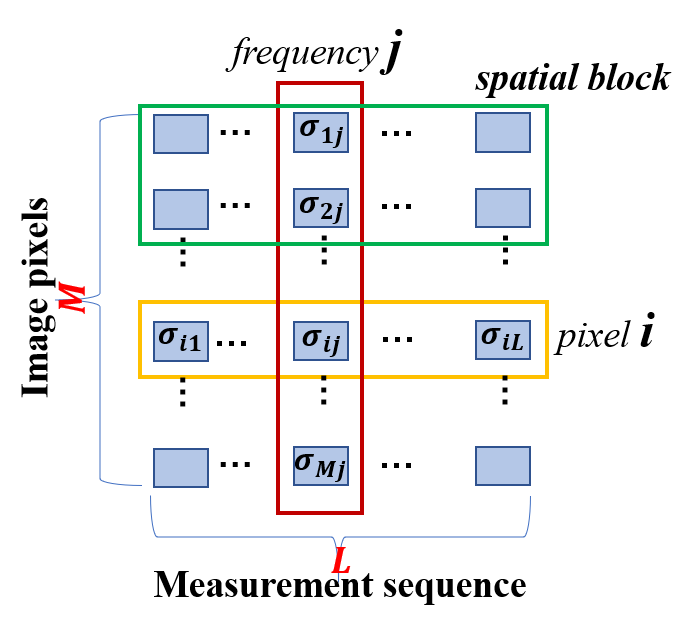}}
	\subfigure[Embedding block structure $\mathbf{A}$]{\includegraphics[width =1.5 in]{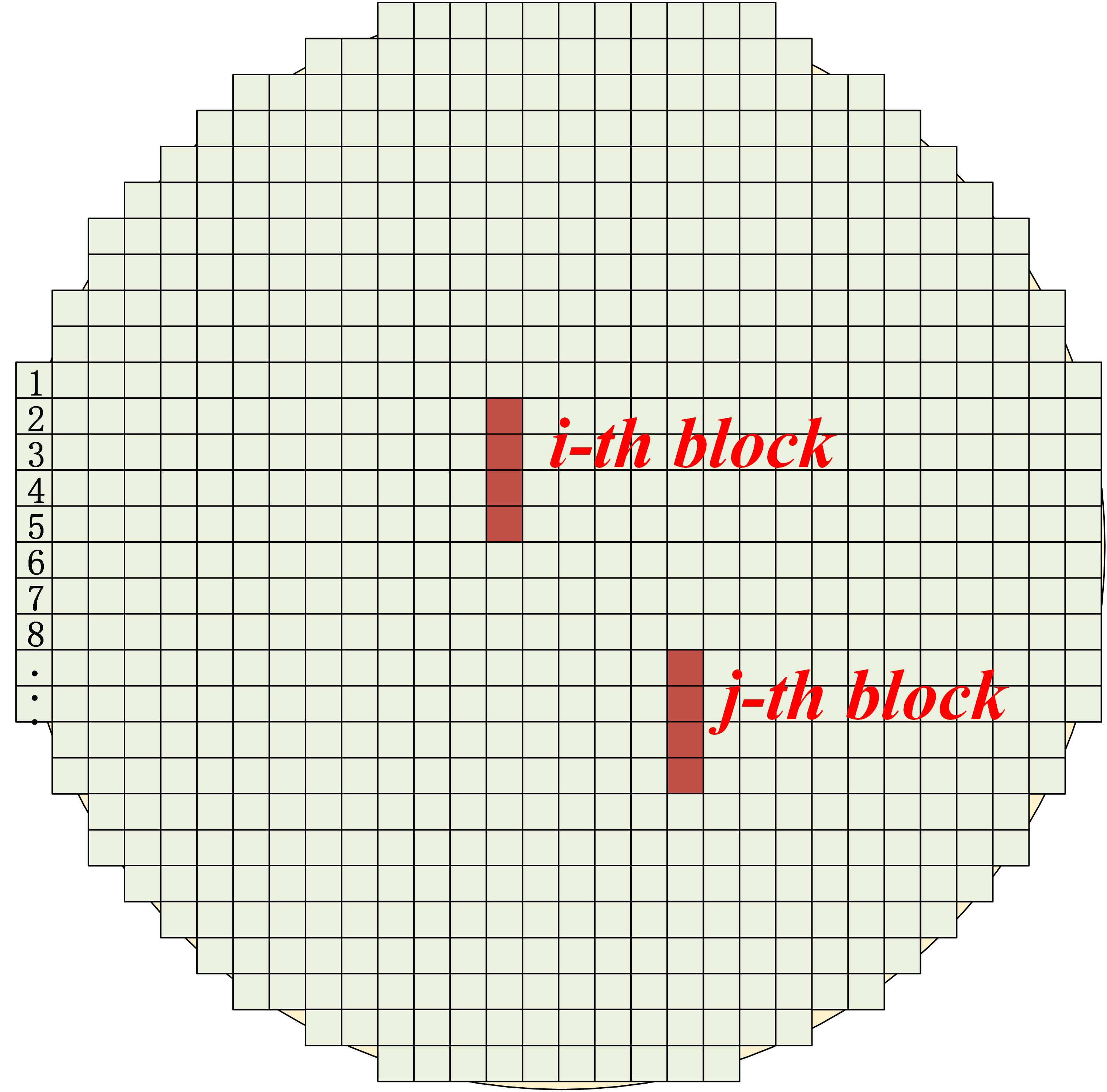}}
	\caption{Schematic illustration of MMV model}
	\label{fig:MMV}
\end{figure}

To utilize the correlation of mfEMT images, we consider the MMV model for mfEMT image reconstruction, which is given by
\begin{equation}
	\mathbf{Y}=\mathbf{J}\boldsymbol{\sigma}+\mathbf{V}
	\label{eq:MMV1}
\end{equation}
where $\mathbf{Y} = [ \mathbf{y}_1, \mathbf{y}_2, \dots \mathbf{y}_L  ] \in \mathbb{R}^{N\times L}$ is the multi-frequency  measurement matrix. $\boldsymbol{\sigma} = [ \boldsymbol{\sigma}_1, \boldsymbol{\sigma}_2, \dots \boldsymbol{\sigma}_L  ] \in \mathbb{R}^{M\times L}$ is the  corresponding conductivity distribution at $L$ frequencies.  As illustrated in Fig. \ref{fig:MMV}(a), each column of $\boldsymbol{\sigma}$ is associated with a measurement at given frequency and each row represents a pixel in the image.  $\mathbf{V} = [ \mathbf{v}_1, \mathbf{v}_2, \dots \mathbf{v}_L  ]\in \mathbb{R}^{N\times L}$ denotes the unknown noise vectors at multiple measurements.

We first make a  block sparsity assumption about the spatial conductivity. Without requiring any prior knowledge about the block distribution pattern, we consider the general case that the blocks have equal size of $h$ and overlap with each other \cite{liu2018image}. As shown in Fig. \ref{fig:MMV}(b),  the block structure $\mathbf{A}$ is embedded into the sensitivity matrix
\begin{equation}
	\boldsymbol{\sigma} \triangleq \mathbf{A} \mathbf{x} \triangleq\left[\mathbf{A}_{1}, \ldots, \mathbf{A}_{g}\right]\left[\mathbf{x}_{1}^{\mathrm{T}}, \ldots, \mathbf{x}_{g}^{\mathrm{T}}\right]^{\mathrm{T}}
	\label{eq:MMV2}
\end{equation}
where $g = M-h+1$ is the total number of blocks, and $\forall i=1, 2, \dots g$, $\mathbf{A}_{i} \triangleq\left[\mathbf{0}_{(i-1) \times h}^{\mathrm{T}}, \mathbf{I}_{h \times h}^{\mathrm{T}}, \mathbf{0}_{(M-i-h+1) \times h}^{\mathrm{T}}\right]^{\mathrm{T}}\in \mathbb{R}^{M \times h}$ denotes the $i$-th block structure and $\mathbf{x}_{i}= \left[x_{i}, \ldots, x_{i+h-1}\right]^{\mathrm{T}} \in \mathbb{R}^{h \times 1}$ denotes the weights of this block.

The spatial block-sparse underlying model in Eq. (\ref{eq:MMV1}) is expressed as 
\begin{equation}
	\mathbf{Y}=\mathbf{J}\boldsymbol{\sigma}+\mathbf{V}=\mathbf{J}\mathbf{A}\mathbf{x} + \mathbf{V}
	\label{eq:MMV3}
\end{equation}

To solve this MMV problem, we transfer the temporally correlated MMV (T-MSBL) method in \cite{zhang2011sparse} to our frequency-correlated case.  Then Eq. (\ref{eq:MMV3}) can be rewritten as
\begin{equation}
	\mathbf{y}_F = \mathbf{D} \mathbf{x}_F + \mathbf{v}_F 
	\label{eq:MMV4}
\end{equation}
by letting $\mathbf{y}_F = \operatorname{vec}\left(\mathbf{Y}^{T}\right)\in \mathbb{R}^{N L \times 1}$, $\mathbf{D} = \mathbf{J} \mathbf{A} \otimes \mathbf{I}_{L}$, $\mathbf{x}_F = \operatorname{vec}\left(\mathbf{X}^{T}\right)\in \mathbb{R}^{(ghL) \times 1}$, $\mathbf{v}_F = \operatorname{vec}\left(\mathbf{V}^{T}\right)\in \mathbb{R}^{NL \times 1}$.
$\otimes$ represents the Kronecker product of the two matrices. $\operatorname{vec}(.)$ denotes the vectorization of the matrix  by stacking its columns into a single column vector.

We assume that the elements in the noise vector $\mathbf{v}_F$ are independent and each has a Gaussian distribution i.e.  $p\left(\mathbf{v}_F\right) \sim \mathcal{N}(0, \gamma_0\mathbf{I})$.
To exploit the spatial block sparsity and frequency information simultaneously, we design a prior of the weights $\mathbf{x}_F$ using a zeromean Gaussian distribution
\begin{equation}
	p\left(\mathbf{x}_F ; \gamma_{i}, \mathbf{B}_{i}, \forall i\right) \sim \mathcal{N}_{x}\left(\mathbf{0}, \mathbf{\Sigma_{0}}\right)
	\label{eq:sbl1}
\end{equation}
where $\mathbf{\Sigma_0}$ is 
\begin{equation}
	\mathbf{\Sigma}_{0}=\left[\begin{array}{cc}{\gamma_{1} \mathbf{B}_{1}} & {\mathbf{0}} \\ {} & {\ddots} \\ {\mathbf{0}} & {\gamma_{g} \mathbf{B}_{g}}\end{array}\right] \in \mathbb{R}^{g h L\times g h L}
	\label{eq:sbl2}
\end{equation}
and $\mathbf{B}_i\in \mathbb{R}^{hL\times hL}$  incorporates the spatial and frequency information into one covariance matrix. To avoid overfitting, it is a good practice to use one positive definite matrix $\mathbf{B}$ to model all the pixel covariance matrices $\mathbf{B}_i, \forall i$. The spatial sparsity is controlled by $\gamma_i, \forall i$. 

With this prior, Eq. (\ref{eq:MMV4}) is solved using Bayesian rule that seeks to optimize the posterior of $p(\mathbf{x}_F|\mathbf{y}_F)$, i.e.
\begin{equation}
	p(\mathbf{x}_F | \mathbf{y}_F ; \left\{\gamma_{0},\left\{\gamma_{i}, \mathbf{B}_{i}\right\}_{i=1}^{g}\right\})=\mathcal{N}\left(\boldsymbol{\mu}_{x}, \mathbf{\Sigma}_{x}\right)
	\label{eq:sbl3}
\end{equation}

The final solution is $\mathbf{\hat{x}} = \boldsymbol{\mu}_x$. The hyperparameter estimation of $\boldsymbol{\mu}_x, \mathbf{\Sigma}_x $ adopts the T-MSBL method introduced in \cite{zhang2011sparse}. It is an efficient Bayesian method  in low dimensional space, even with the increase of measurements.

\section{ Results and Discussions}
\label{sec:Experiment}

\subsection{System Evaluation}

\begin{figure}[tbp]
	\centering
	\includegraphics[width = 3.2 in]{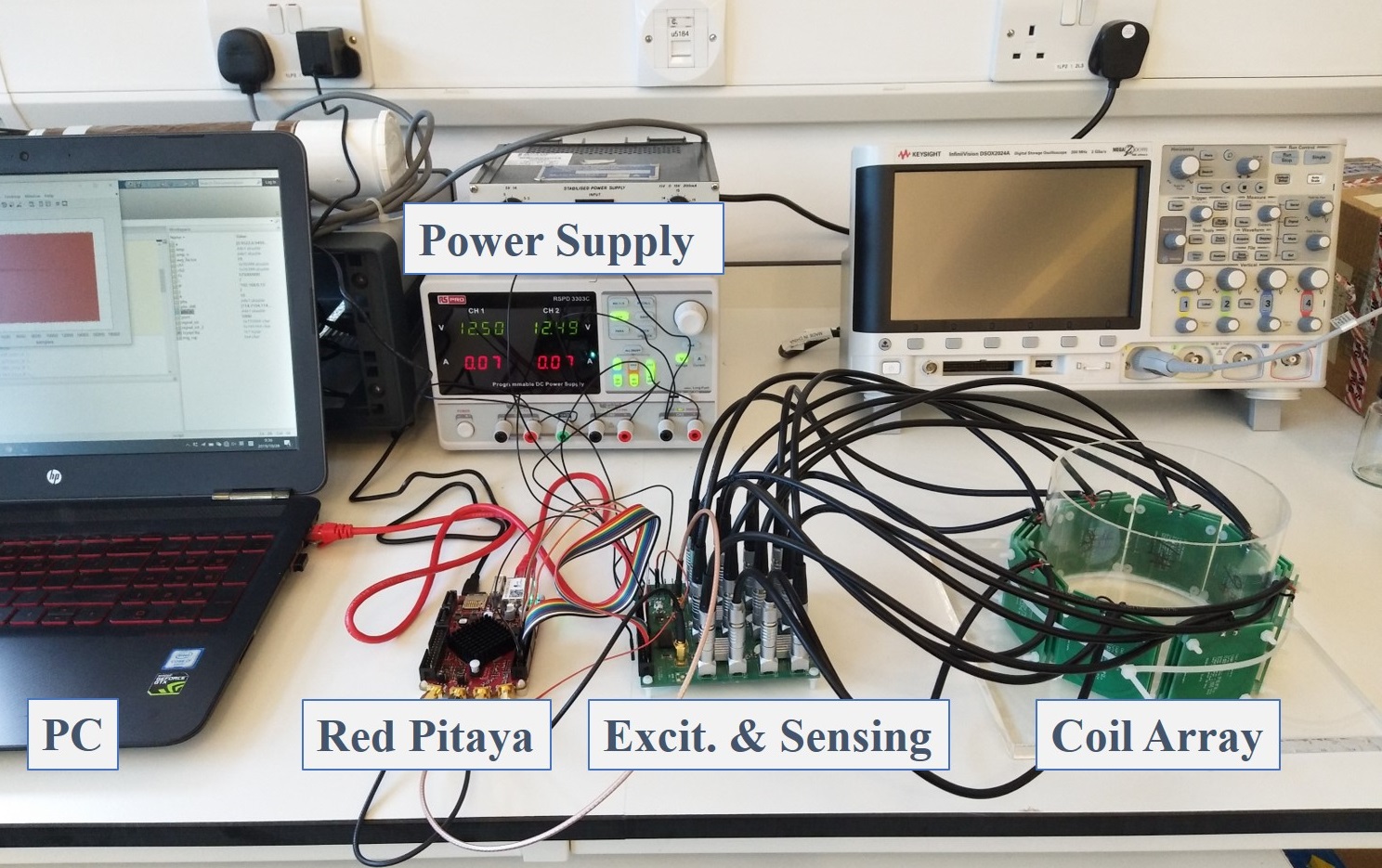}
	\caption{Experimental setup.}
	\label{fig:System2}
\end{figure}

\begin{figure}[tbp]
	\centering
	\includegraphics[width = 3.5 in]{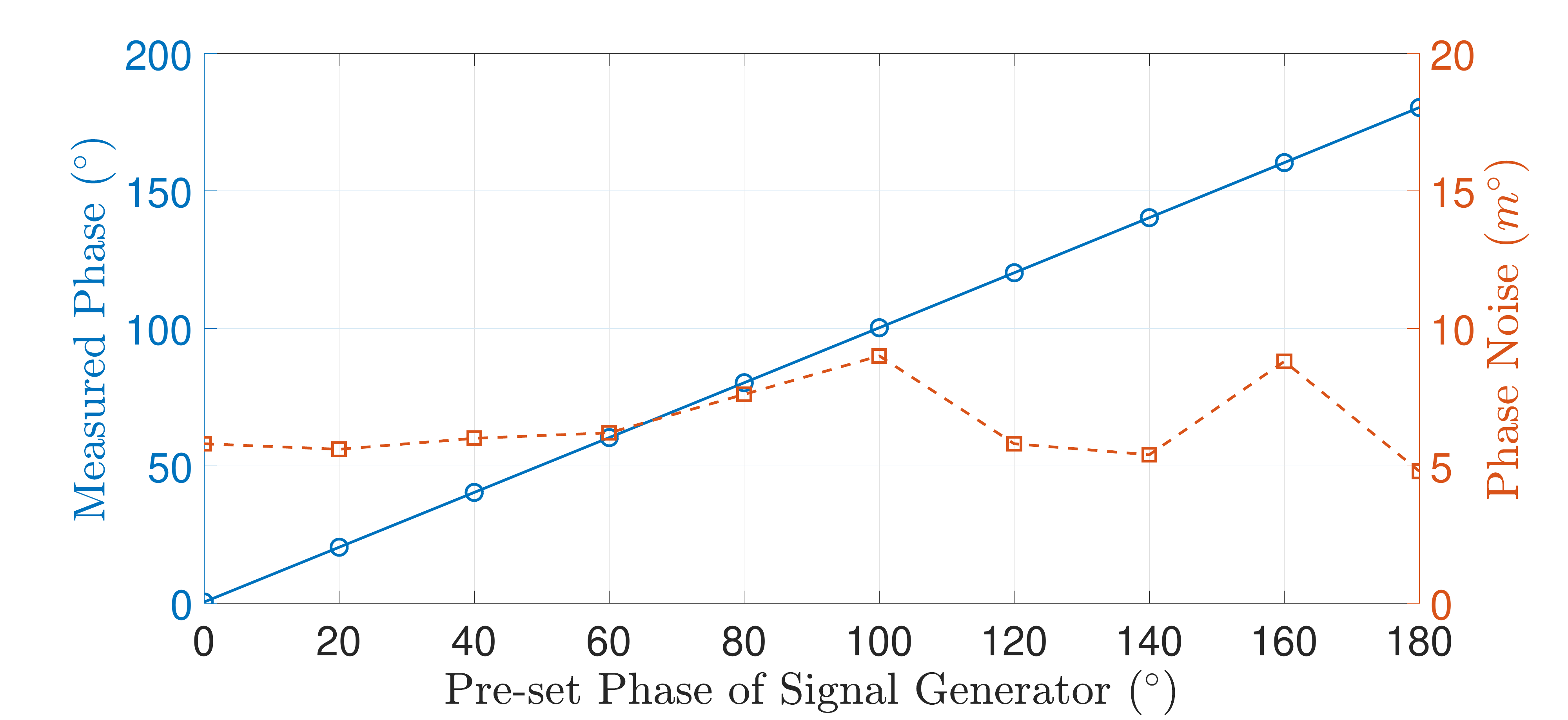}
	\caption{Phase measurement linearity and noise.}
	\label{fig:SensorLine}
\end{figure}

Fig. \ref{fig:System2} shows the setup of the mfEMT system which consists of a coil array with a diameter of 120 mm, a signal generation and data acquisition module based on Red Pitaya,  a computer for image reconstruction, and a circuit that incorporates multiplexer, excitation , and sensing electronics.

To evaluate the phase measurement accuracy of Red Pitaya,  we used the Agilent 33500 signal generator to produce dual-channel signals with predefined phase differences. The signals with an amplitude of 500 mVpp at 6.25 MHz are sampled by Red Pitaya at 125 MSPS.  The FFT data length is 16384 (maximum capacity of red pitaya buffer) and the averaging factor is 20. Fig. \ref{fig:SensorLine} illustrates that the phase noise of Red Pitaya is lower than $10\ m^\circ$ and the linearity is persistent over the whole measurement range. It should be noticed that the phase noise  depends on the signal amplitude as the quantization error of ADC is fixed.

\begin{figure}[tbp]
	\centering
	\includegraphics[width = 3.5in]{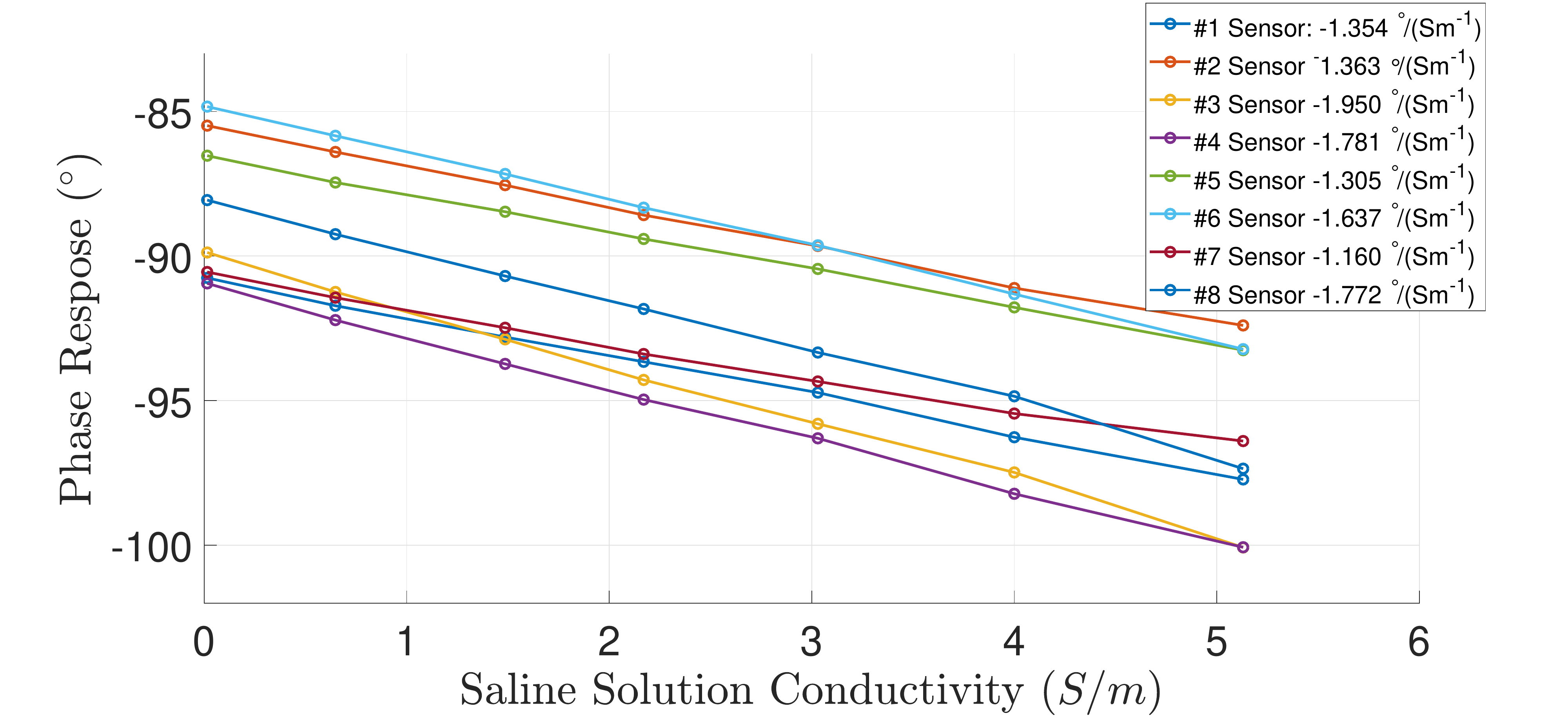}
	\caption{Sensitivity calibration of gradiometer coil .}
	\label{fig:SensCal}
\end{figure}

The sensitivity of 8 channels is calibrated using a sequence of NaCl solution with conductivity ranges from 0.01 S/m to 5.13 S/m. The NaCl solution in plastic bottles was placed closely to the coil which is driven by a 6.25 MHz sinusoidal signal.  As shown in Fig. \ref{fig:SensCal}, the phase sensitivity of the gradiometer coils ranges from  $-1.1^{\circ}/(S\cdot m^{-1})$ to  $-1.9^{\circ}/(S\cdot m^{-1})$. The phase response is linearly dependent on conductivity. Due to the manufacturing inconsistency of coils, the sensitivity of 8 channels should be calibrated before imaging.

\subsection{Image Reconstruction Results}

\begin{figure}[tbp]
	\centering
	\includegraphics[width = 3.5 in]{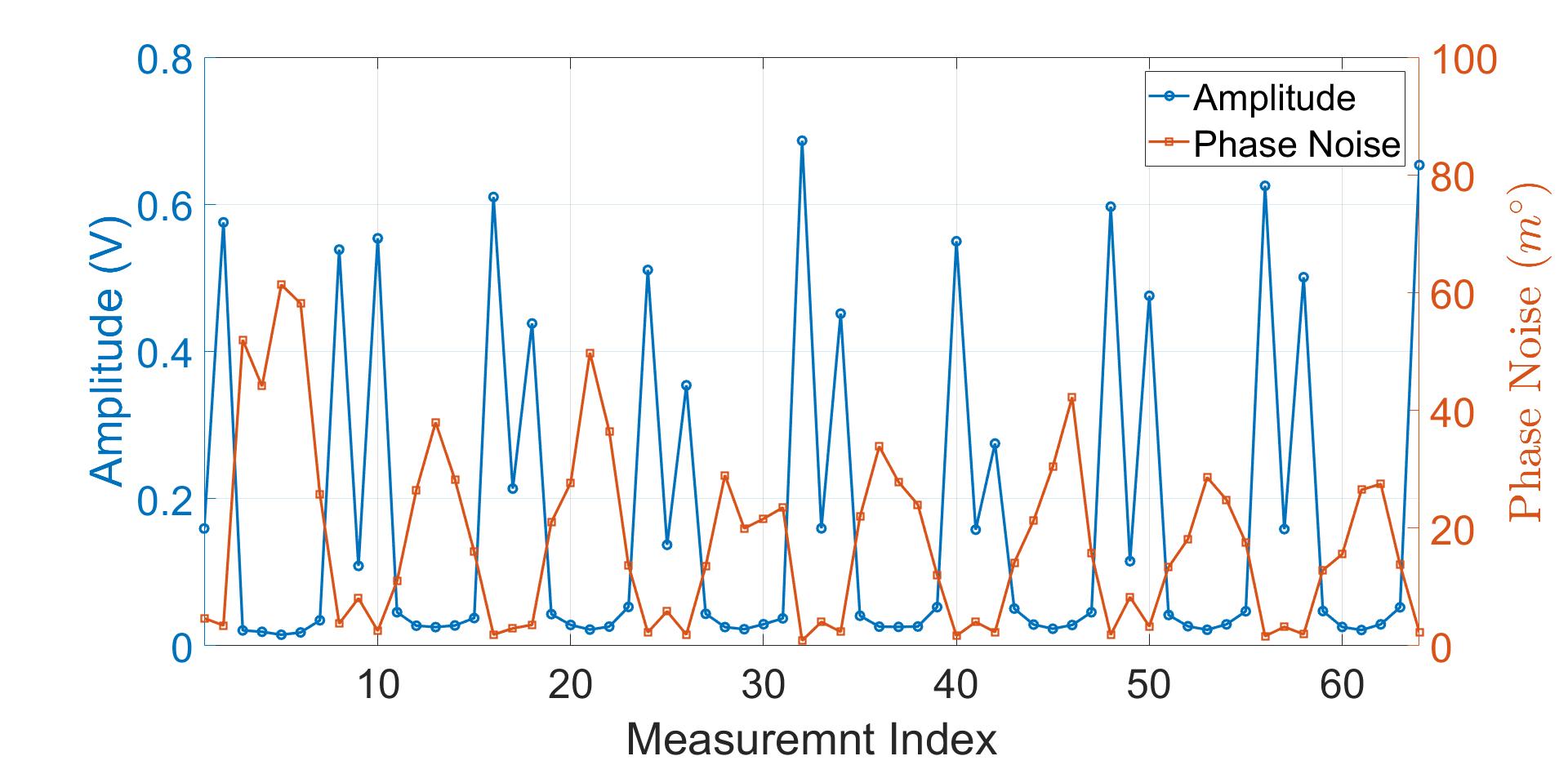}
	\caption{One frame of signal amplitude and phase noise.}
	\label{fig:AmpFrame}
\end{figure}

\begin{table*}[htb]
	\caption{mfEMT Phantoms(Empty, Phantom 1, Phantom 2, Phantom 3)}
	\begin{spacing}{1.5}
		\centering
		\begin{tabular}{|c|c|c|c|} 
			\hline 
			\includegraphics[width=1.1 in]{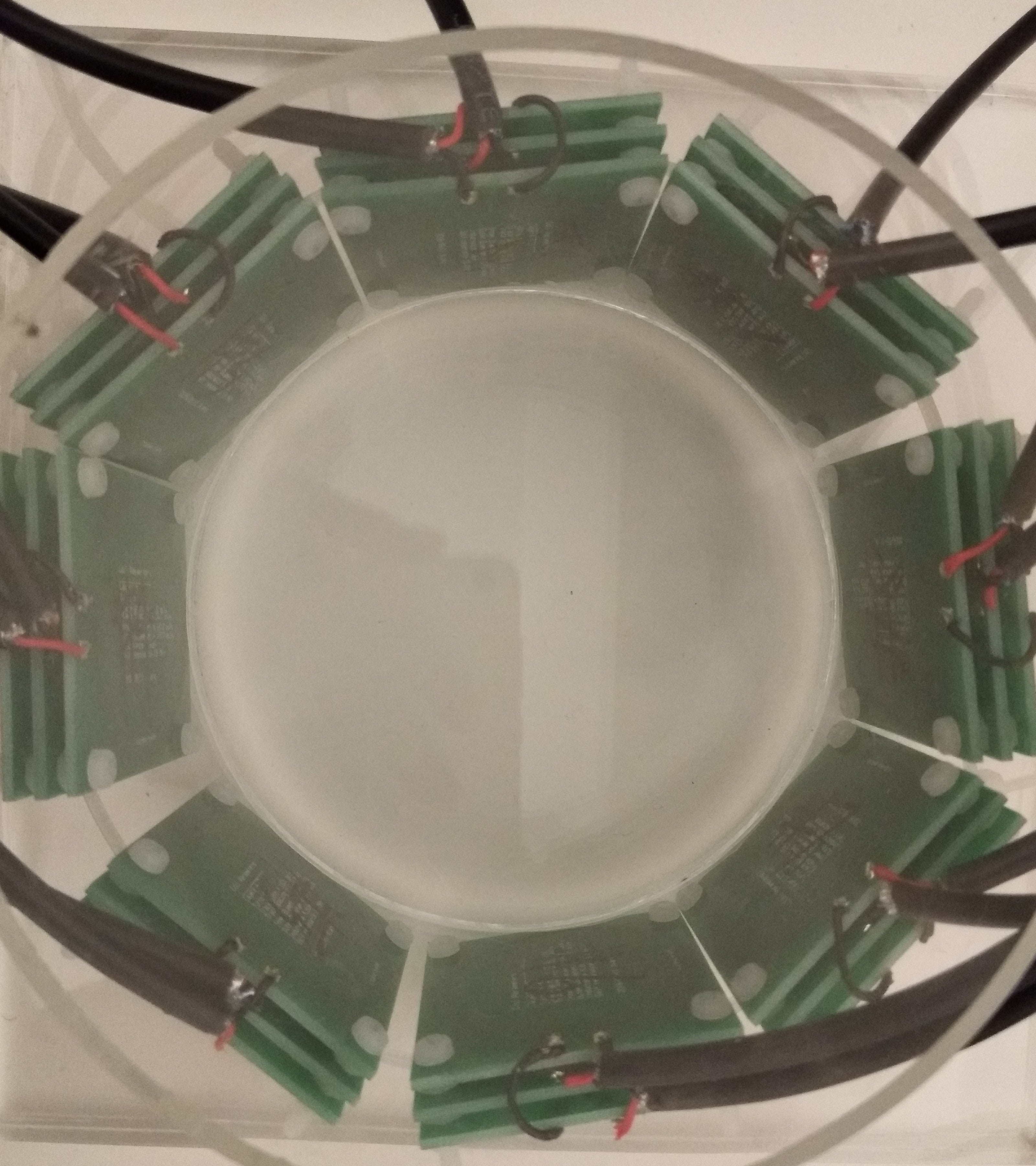} & 
			\includegraphics[width=1.2 in]{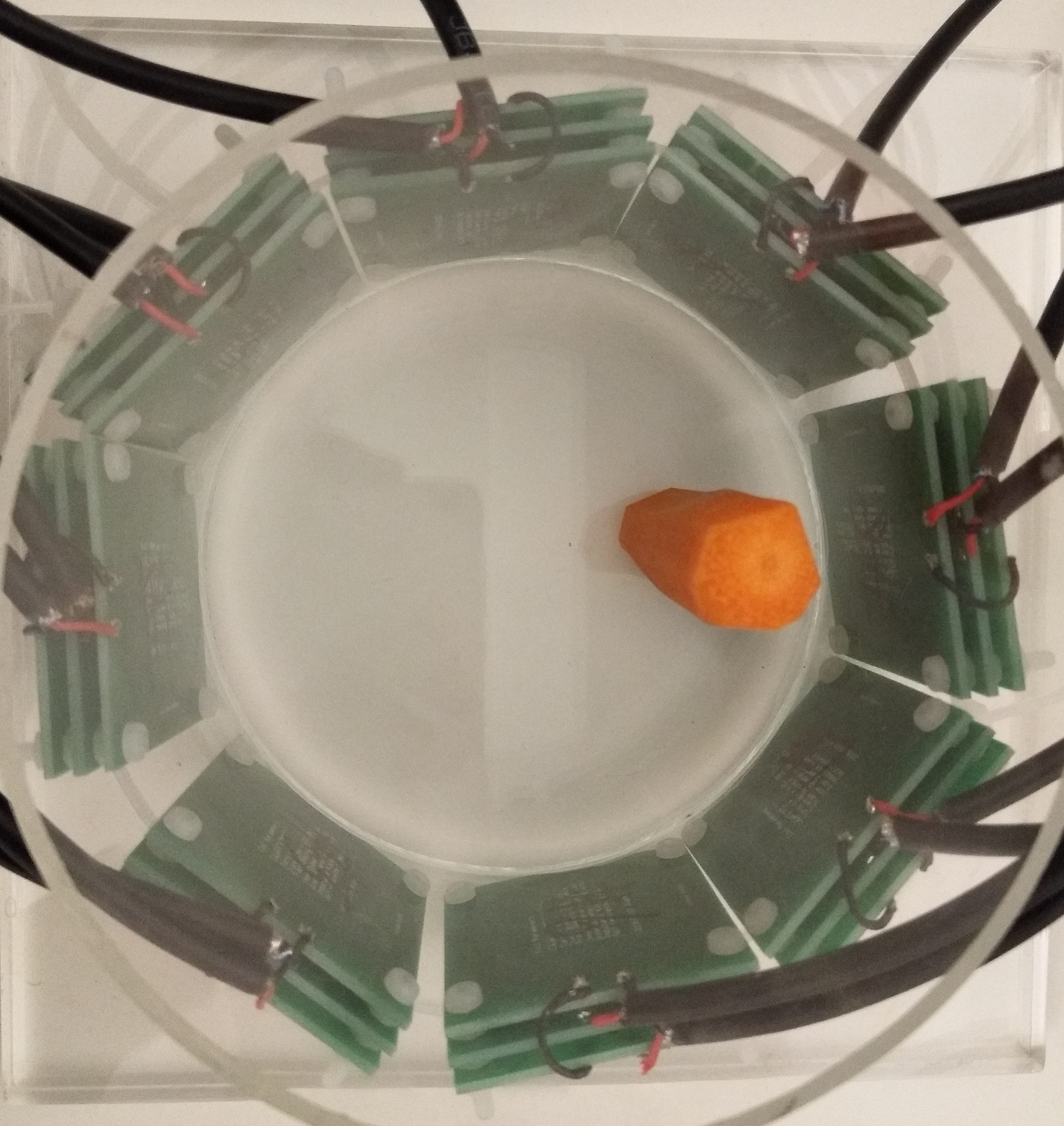} & 
			\includegraphics[width=1.2 in]{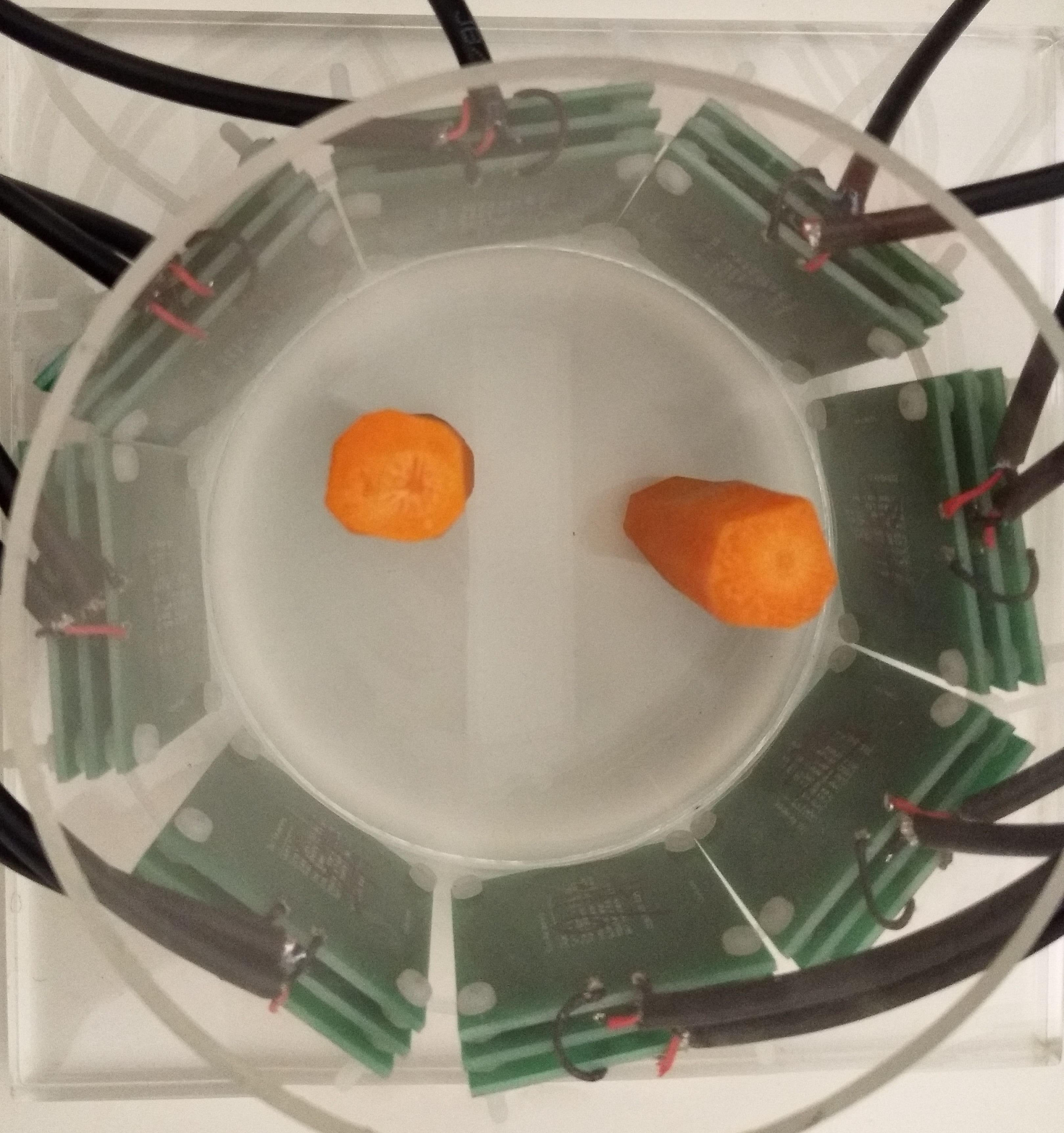} & 
			\includegraphics[width=1.2 in]{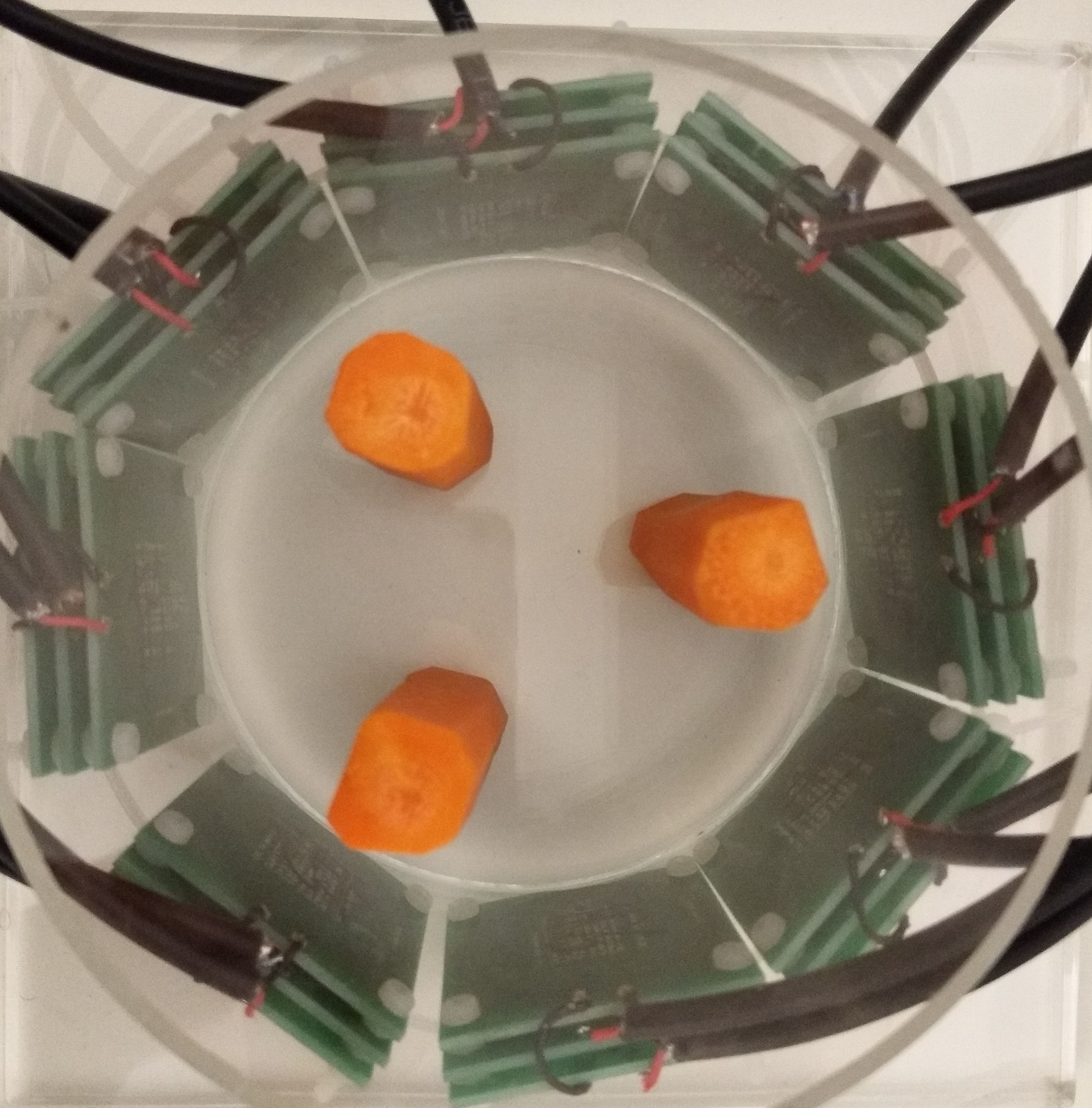}  \\
			\hline  
		\end{tabular}
	\end{spacing}
	\label{tab:TMSBLPhantom1}
\end{table*}

Three test phantoms were designed with carrot cylinders (see Table \ref{tab:TMSBLPhantom1}). From left to right, it is empty, phantom 1 , phantom 2  and phantom 3. Four excitation frequencies of mfEMT in the test are $[f_1,\ f_2,\ f_3,\ f_4] = [0.125,\ 0.25,\ 0.5,\ 1]\times 6.25\ MHz$, which falls in the $\beta$ dispersion.  These frequencies are set to meet full-cycle sampling so as to avoid spectral leakage, since the sampling rate of Red Pitaya is  $125\ MHz$. 

Table \ref{tab:TMSBLPhantom2} illustrates the image reconstruction results of three phantoms with simulation data. The diameter of the cylinders is 10 mm.  The frequency-dependent conductivities of the simulation phantom are $[\sigma_{f_1},\ \sigma_{f_2},\ \sigma_{f_3},\ \sigma_{f_4}] = [0.15,\ 0.25,\ 0.45,\ 0.7]S/m$. These parameters are set according to the conductivity spectra of some agricultural produces in \cite{o2015non}.  In order to simulate the background noise of the real system, same quantity of additional white noise is added to each measurement.  
The reference frame is $f_1$. The three columns on the left show the image reconstruction results of the dual frequency imaging, which is solved by a SMV-based Bayesian method in \cite{wipf2007empirical}. At $f_2$ and $f_3$, the conductivity change relative to $f_1$ is small, in order words, the SNR is small. Therefore, the images are corrupted with strong artifacts. At $f_4$, as the conductivity increases, the image restores to a quite satisfying level. The three columns on the right are the MMV solutions solved by the proposed algorithm in Section \ref{sec:MMV}.  The most intuitive finding is that at $f_2$ and $f_3$ where the SNR is small, the images are reconstructed with little artifacts. 

It should be noticed that the reconstructed conductivity of objects with the MMV model has the same trend as true conductivity spectra. For example, the average reconstructed conductivity of phantom 1 in the object region is $[\hat{\sigma}_{f_2},\ \hat{\sigma}_{f_3},\ \hat{\sigma}_{f_4}] = [0.98, \ 1.46, \ 1.89]$, increasing with frequency. Although these values have not been calibrated with absolute conductivity, it verifies that the improvement of image quality is not just 'averaging the measurements'. The underlying information has been extracted in a noisy environment with the MMV model.

Fig. \ref{fig:AmpFrame} illustrates one frame of experimental data at $6.25\ MHz$ with empty sensing region.  The phase noise depends on the signal amplitude, i.e., larger phase noise corresponds to smaller amplitude.  Apart from the phase errors brought by Red Pitaya, other noise is attributed to the circuits.

Table \ref{tab:TMSBLPhantom3} illustrates the results  with experimental data. The diameter of the cylinders is 17 mm. Due to the inevitable capacitive coupling between coils, additional phase shifts at different frequencies would be caused. Thus, we cannot use one frequency measurement as reference frame and instead, we need to calibrate the phase offset at each frequency point. Results show that the reconstructed objects of dual frequency imaging are hardly observable due to strong background noise and weak signal response cause by the small conductivity of carrots. The ill-posedness of EMT is more severe even than parallel techniques such as  ECT and EIT. In contrast, significant improvements in image quality have been observed with the MMV model (three columns on the right). For phantom 1, the profile and position of the carrot can be seen under three frequencies. For phantom 2 and 3, the objects can be found but unwanted artifacts are becoming more obvious.

\begin{table*}[htb]
	\caption{Image reconstruction results based on simulation data (10-mm cylinders, 120-mm sensing region in diameter)}
	\begin{spacing}{1.5}
		\centering
		\begin{tabular}{|ccc|ccc|} 
			\hline 
			\multicolumn{3}{|c|}{Dual Frequency Imaging with SMV} & \multicolumn{3}{|c|}{mfEMT Imaging with MMV}  \\
			\hline
			$f_2 = 1.5625\ MHz$&$f_3 = 3.125\ MHz$&$f_4 = 6.25\ MHz$&$f_2 =1.5625\ MHz$&$f_3 = 3.125\ MHz$&$f_4 = 6.25\ MHz$ \\
			\includegraphics[width=1 in]{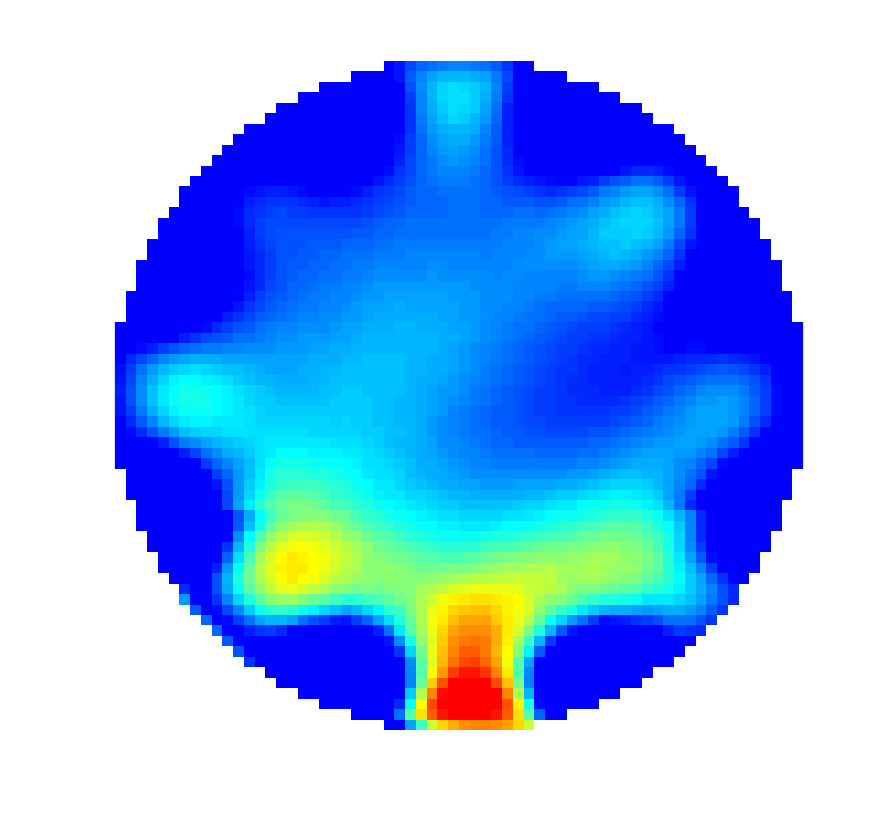} & 
			\includegraphics[width=1 in]{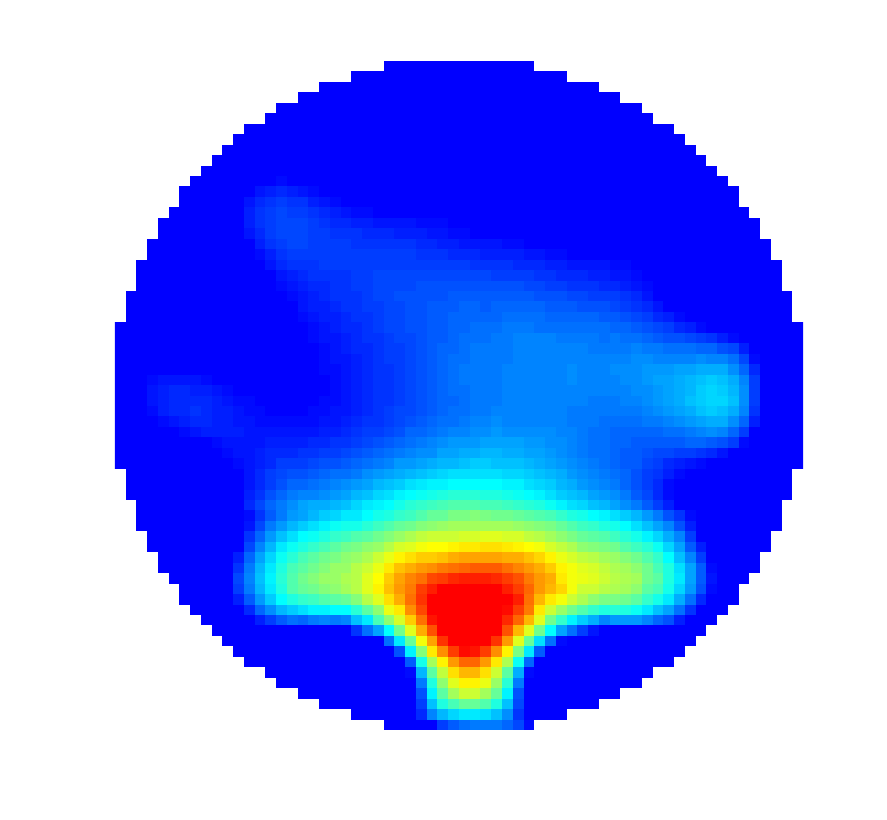} & 
			\includegraphics[width=1 in]{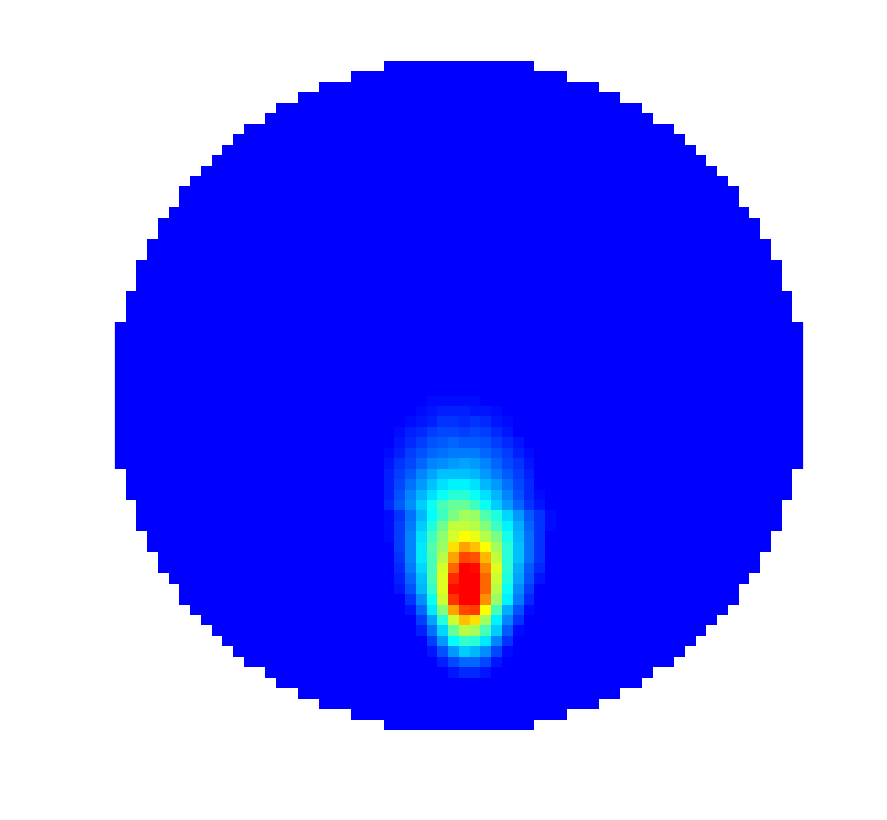} & 
			\includegraphics[width=1 in]{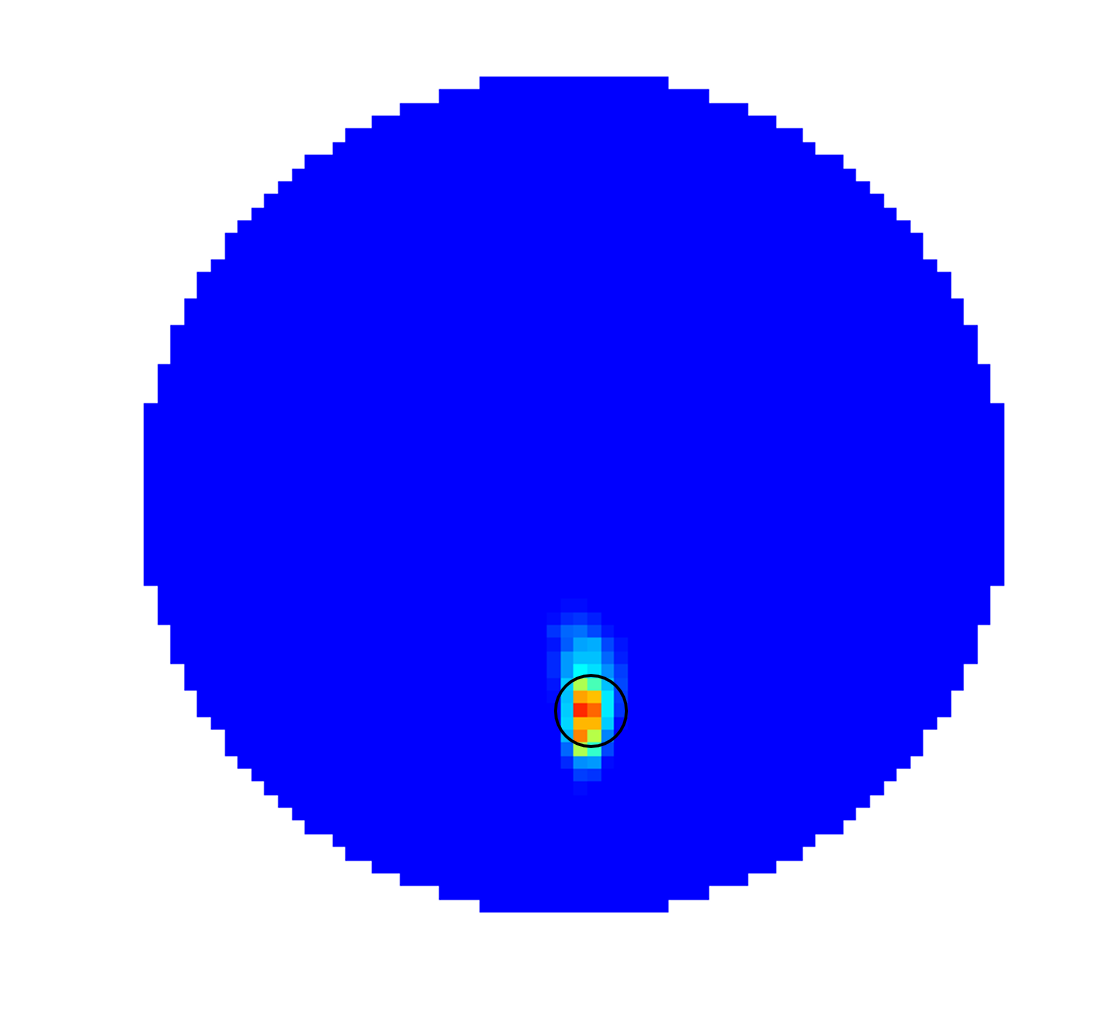} & 
			\includegraphics[width=1 in]{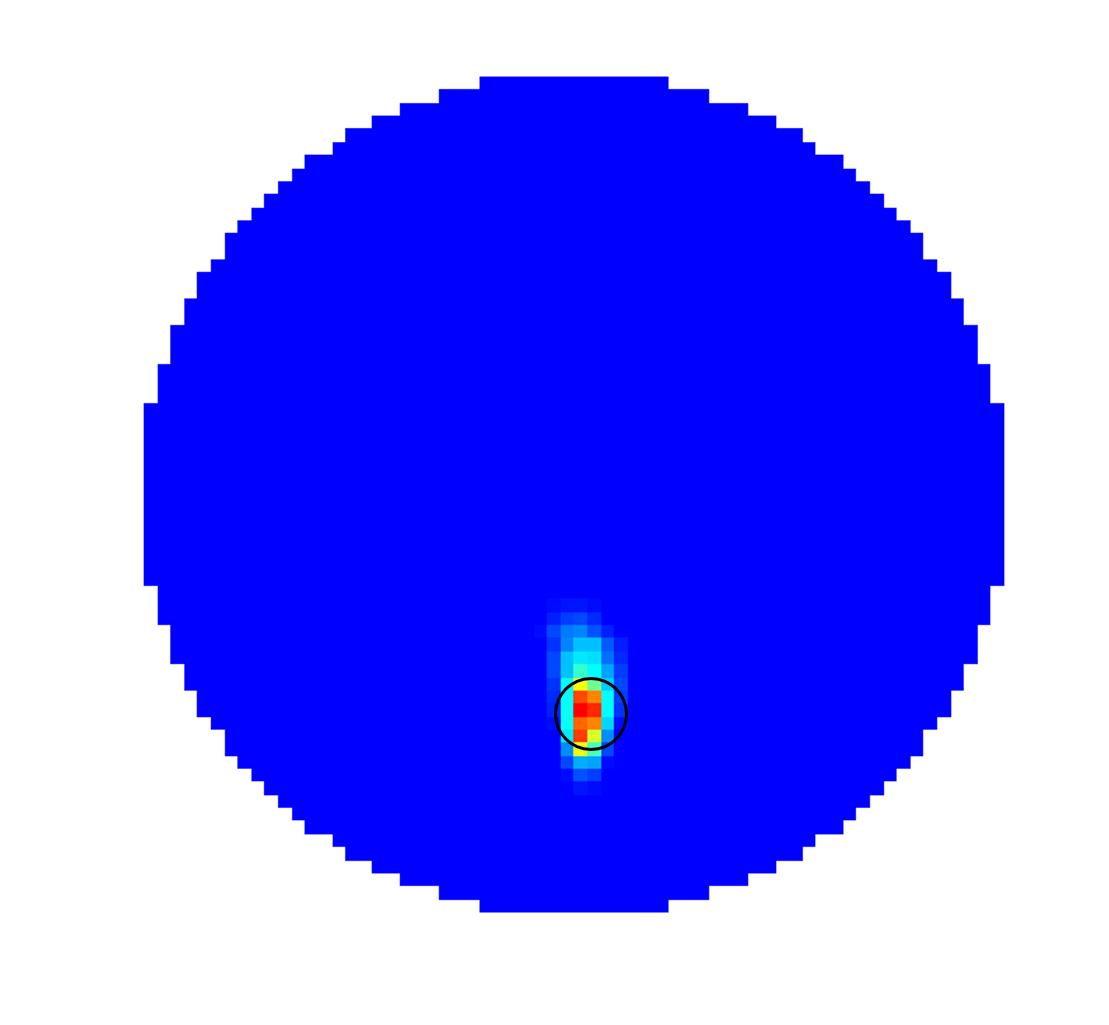} & 
			\includegraphics[width=1.1 in]{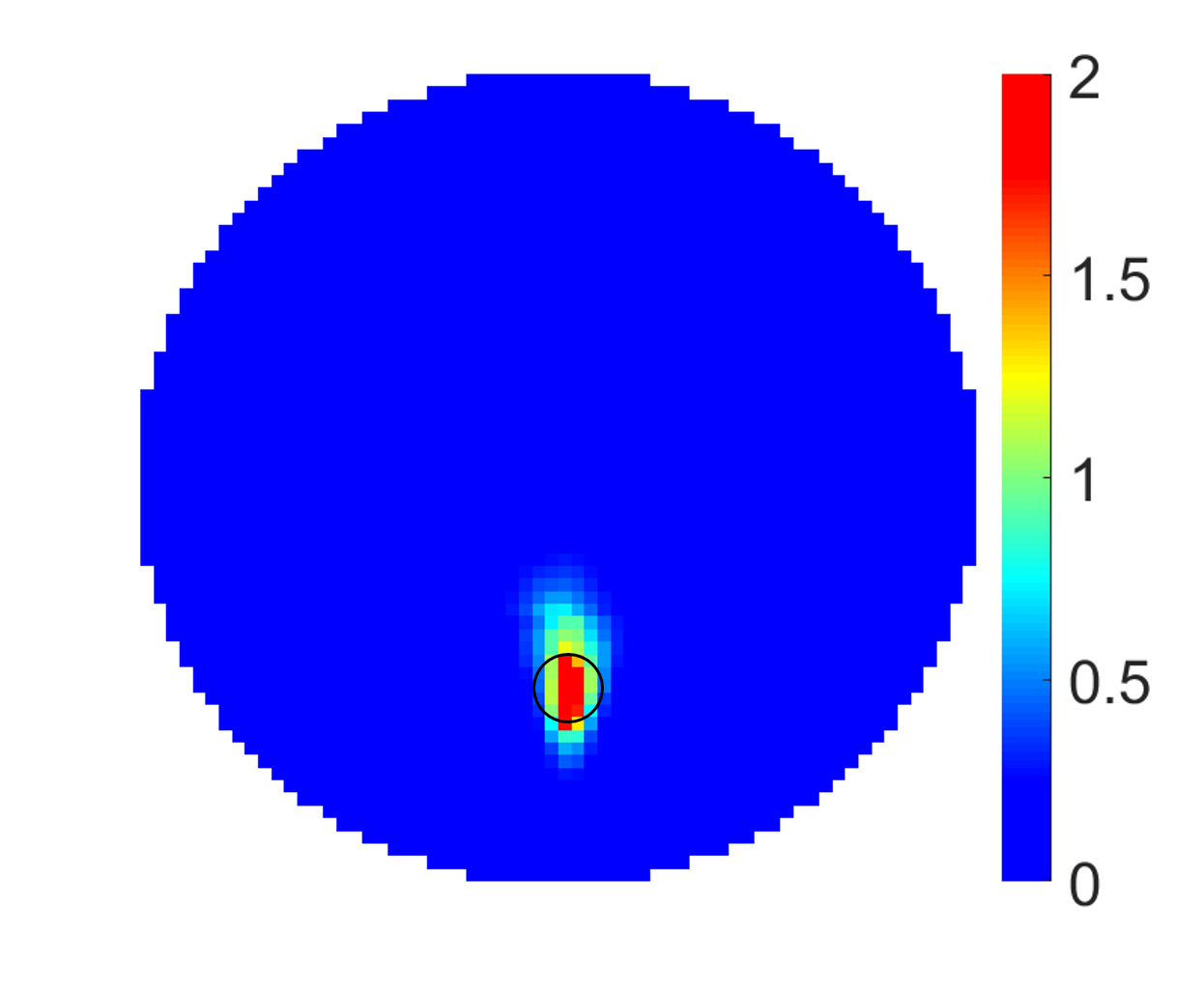} \\
			\includegraphics[width=1 in]{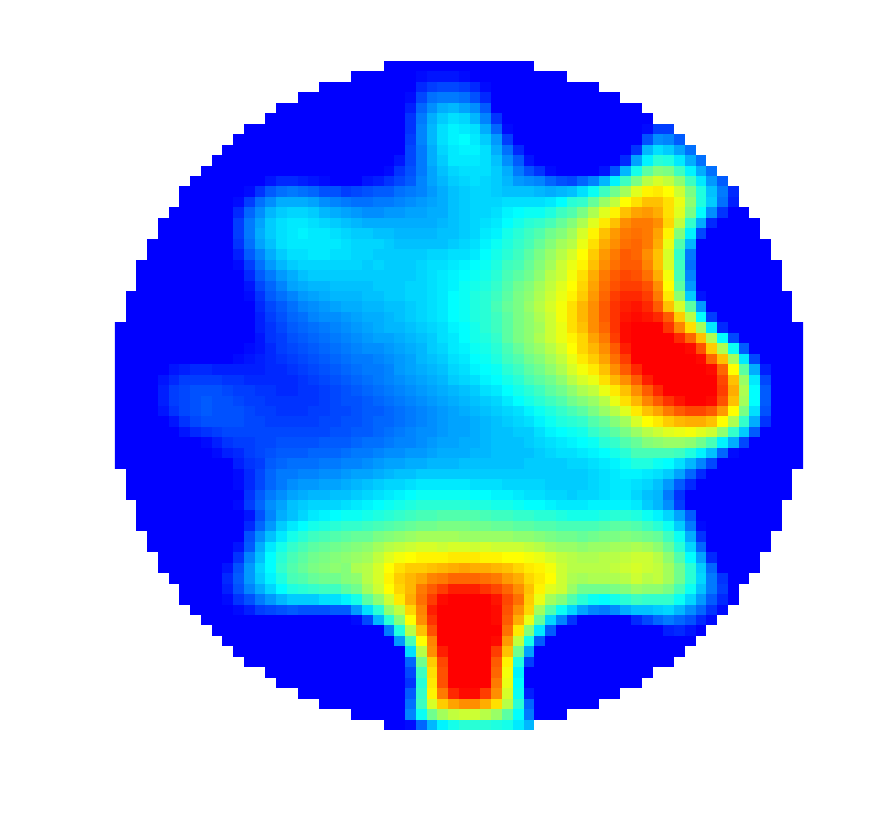} & 
			\includegraphics[width=1 in]{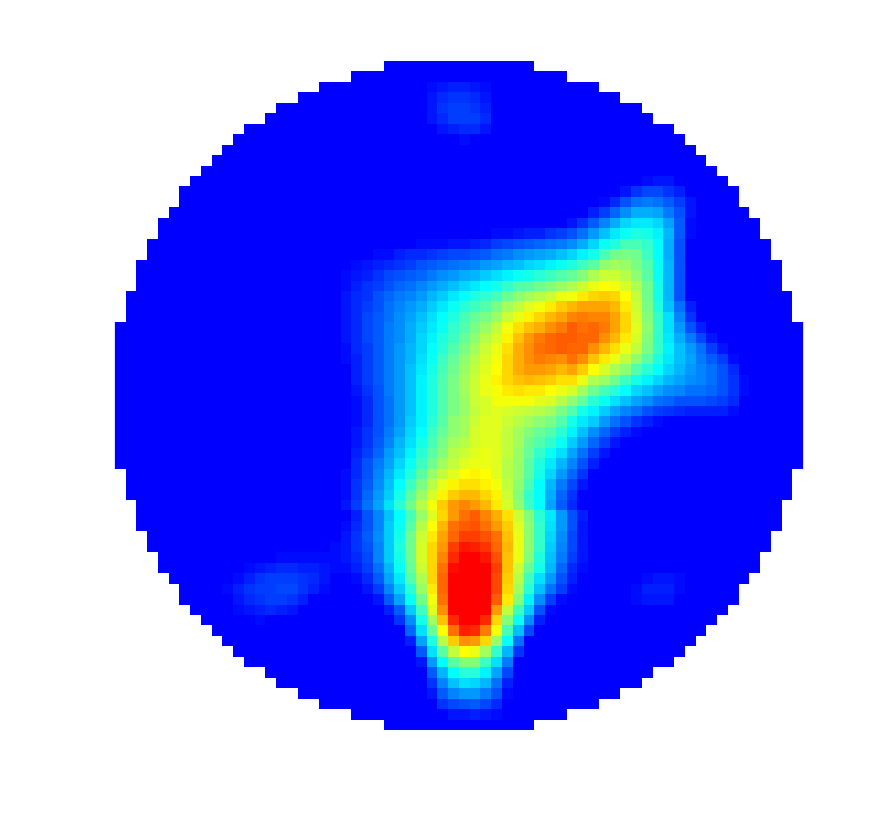} & 
			\includegraphics[width=1 in]{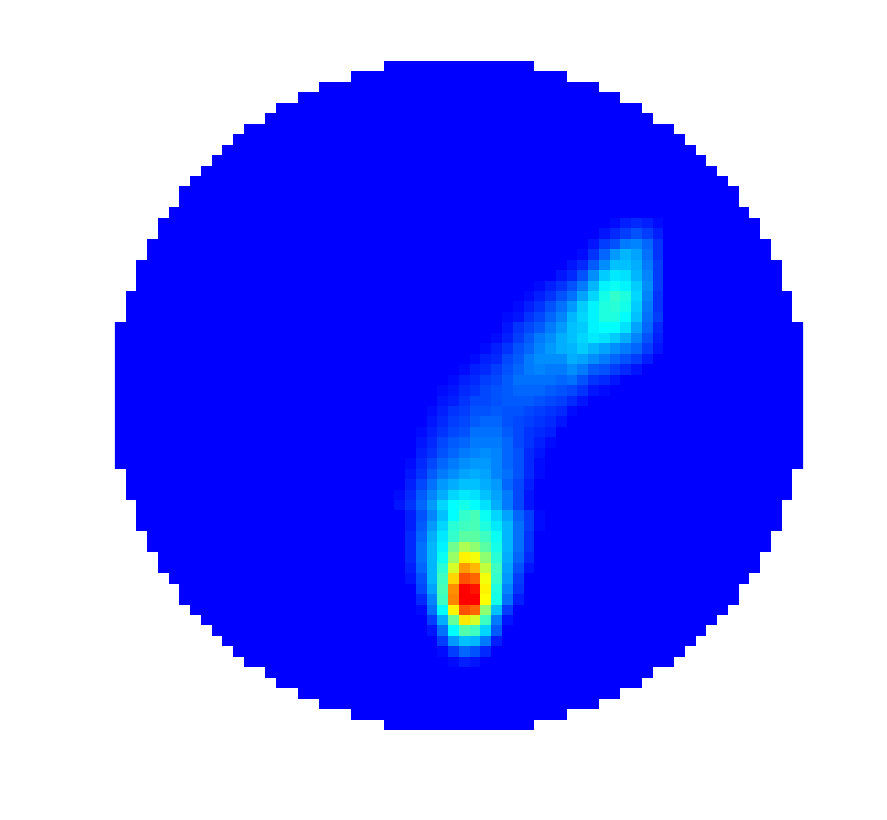} & 
			\includegraphics[width=1 in]{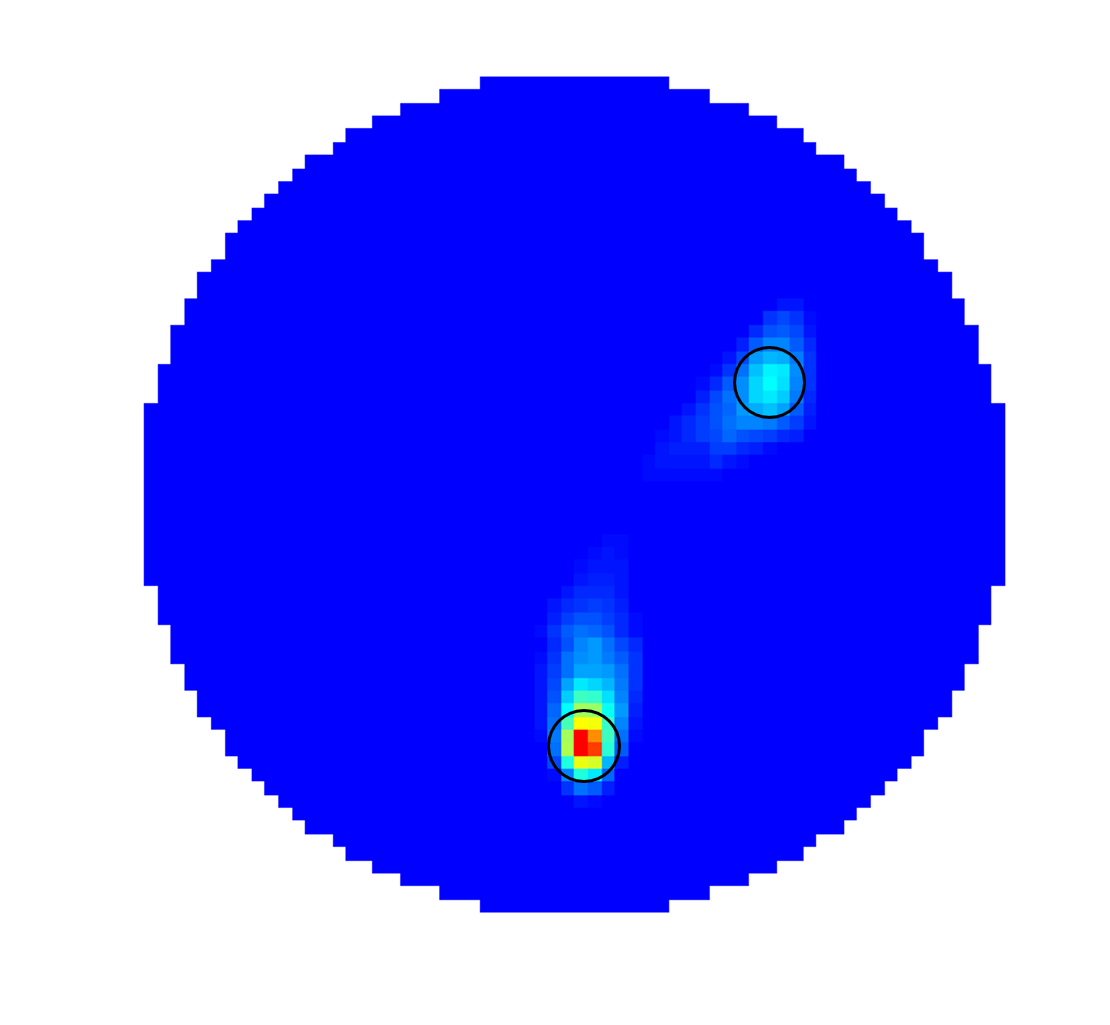} & 
			\includegraphics[width=1 in]{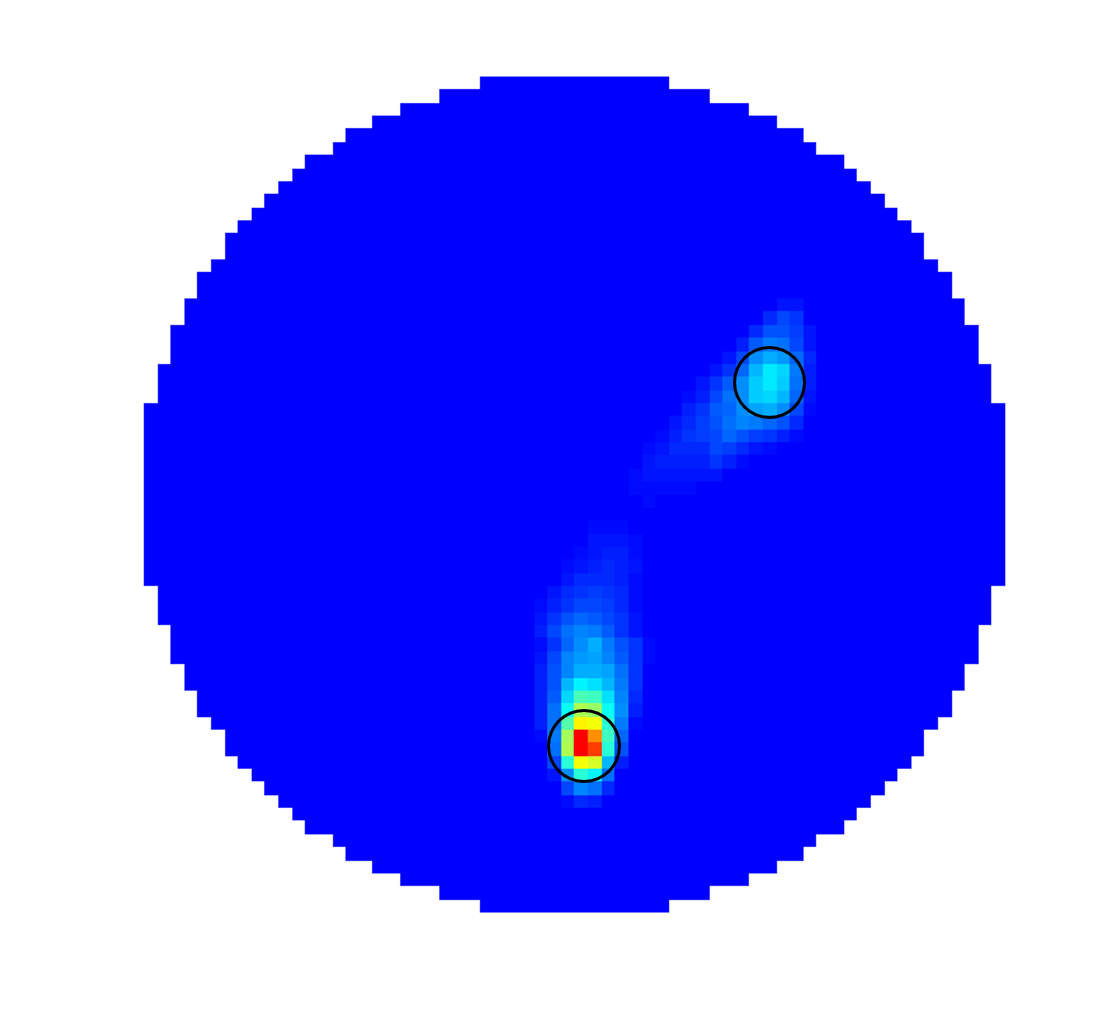} & 
			\includegraphics[width=1.1 in]{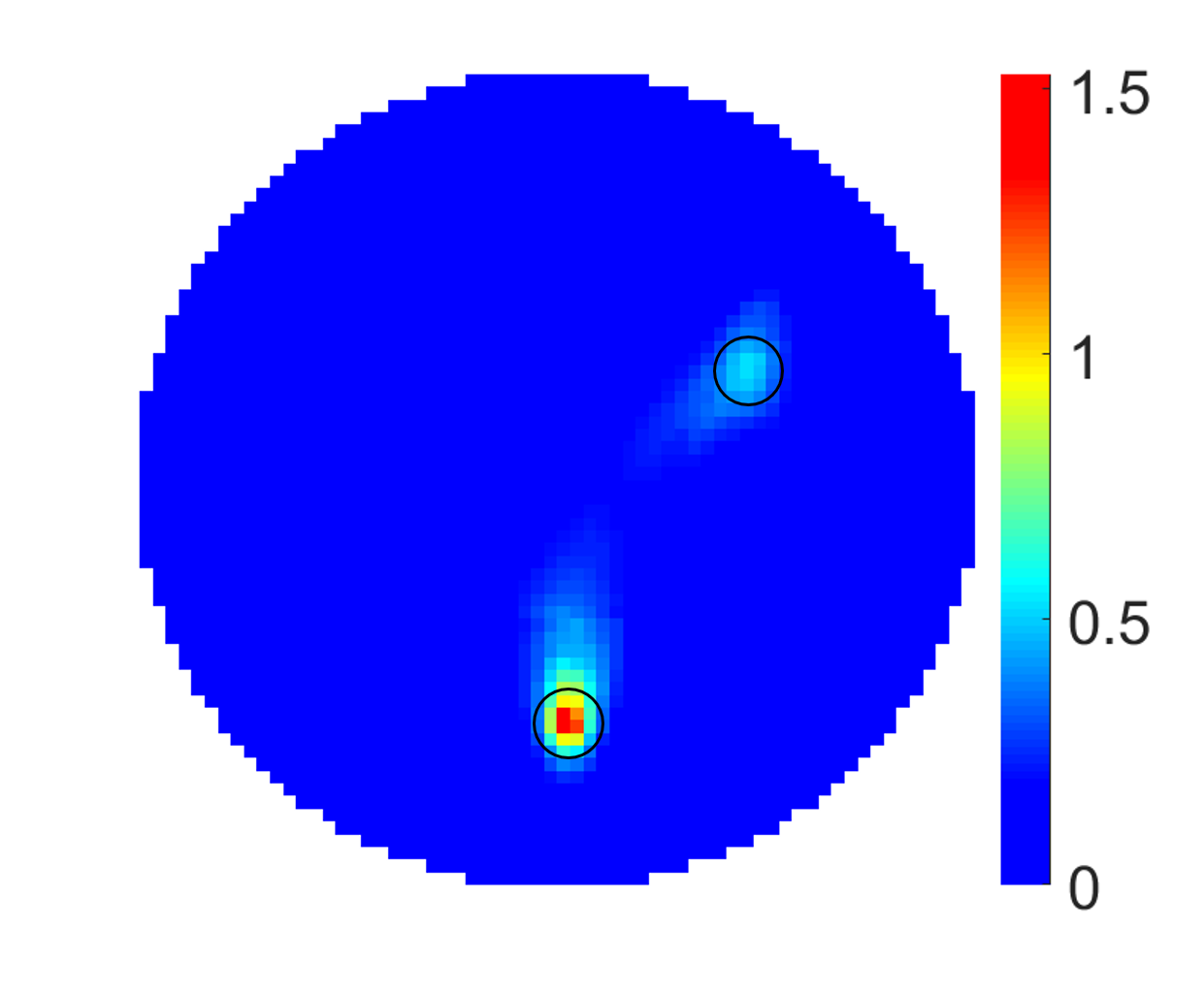} \\
			\includegraphics[width=1 in]{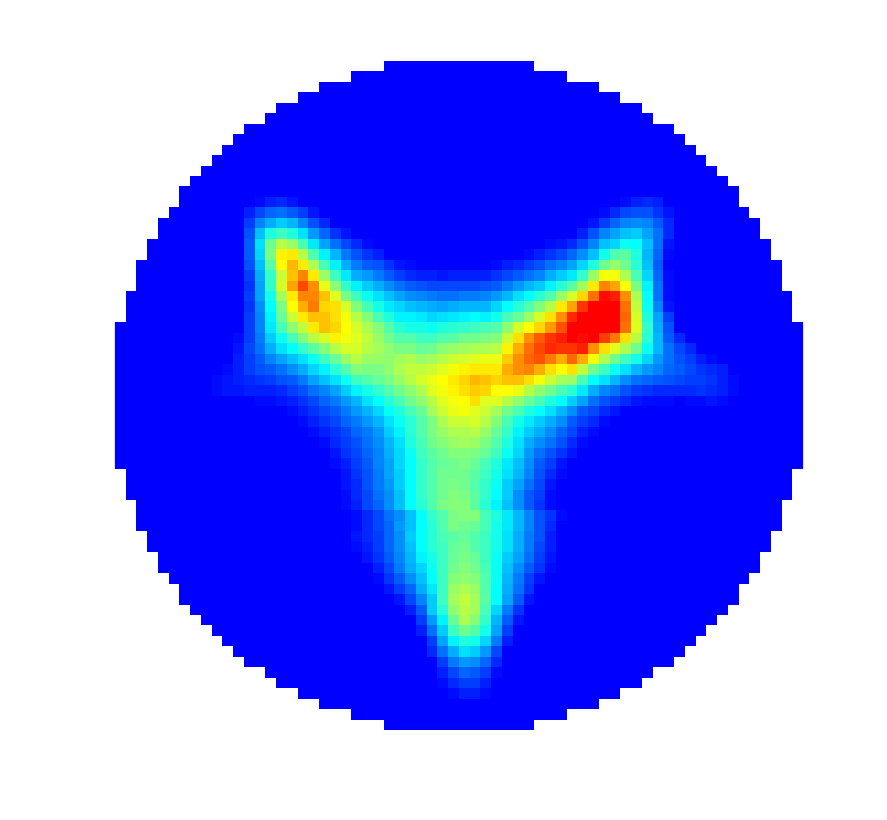} & 
			\includegraphics[width=1 in]{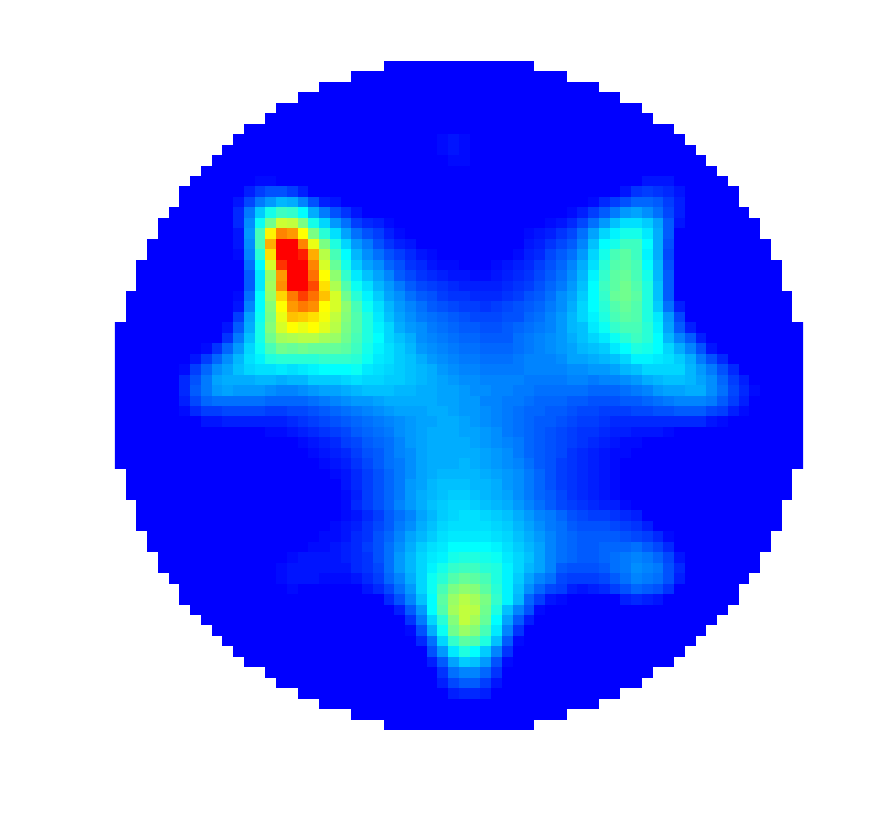} & 
			\includegraphics[width=1 in]{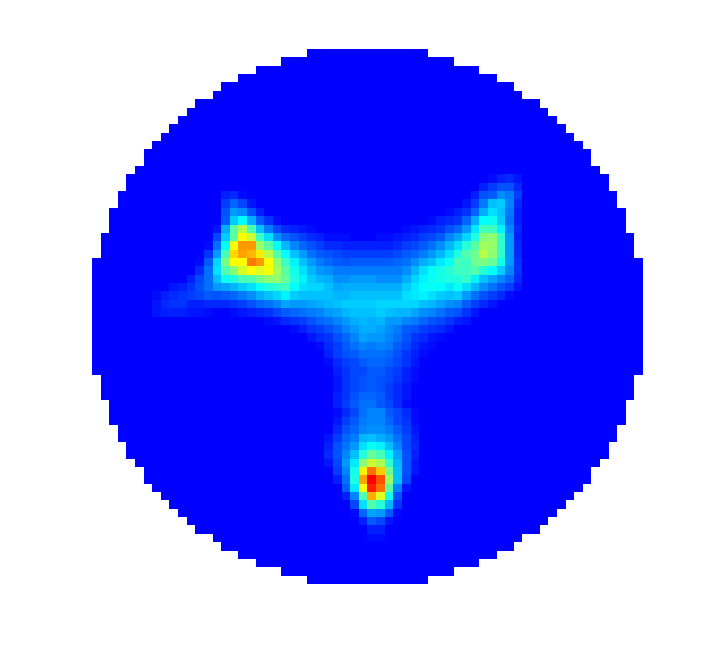} & 
			\includegraphics[width=1 in]{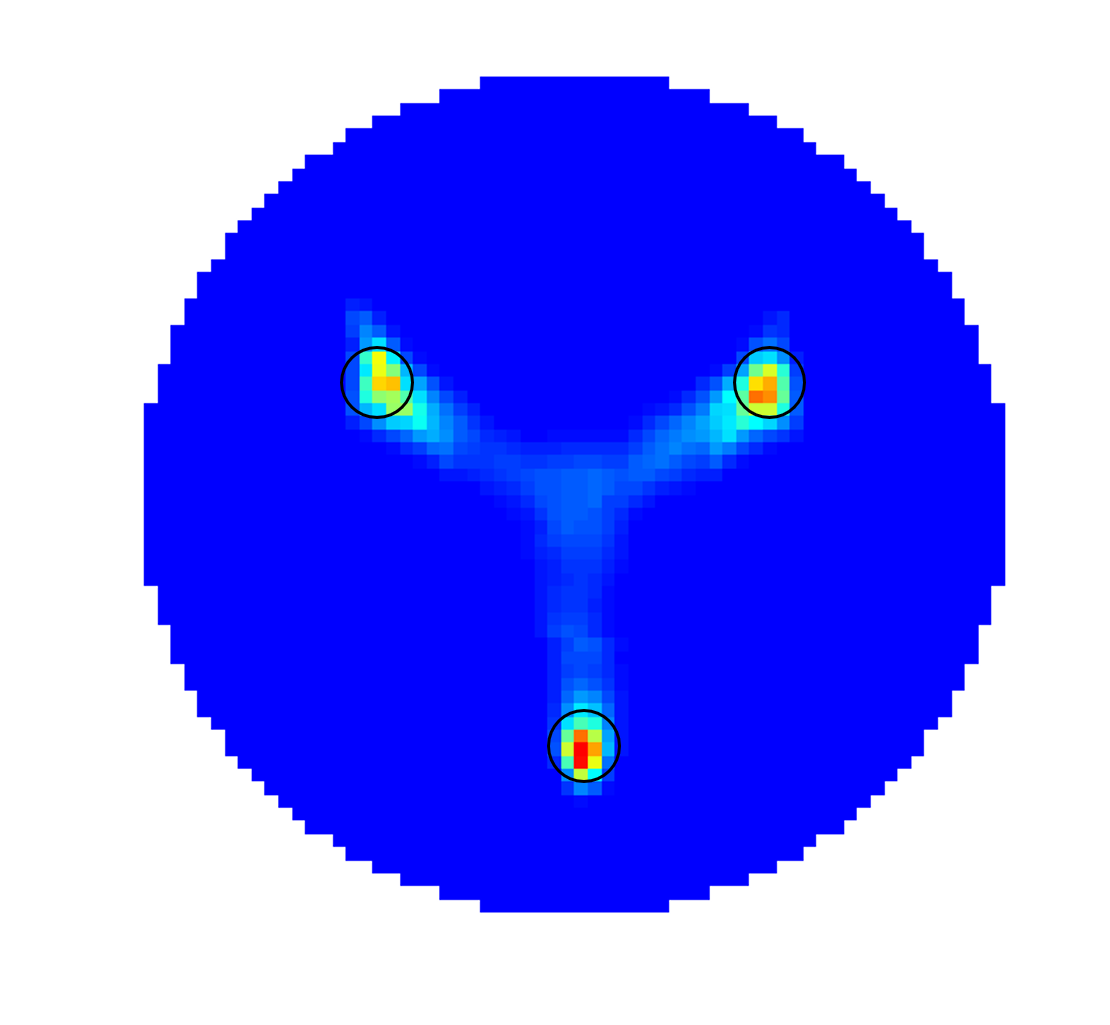} & 
			\includegraphics[width=1 in]{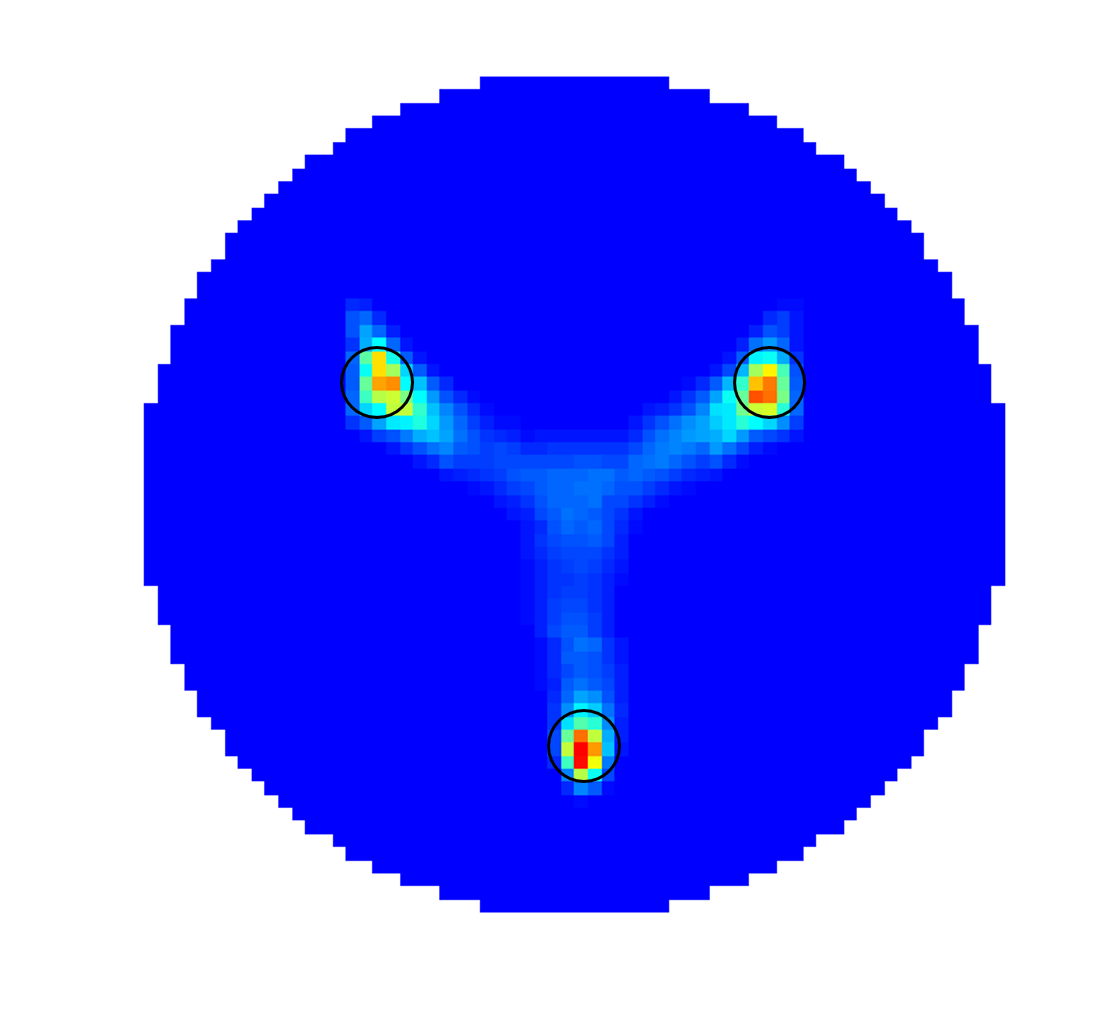} & 
			\includegraphics[width=1.1 in]{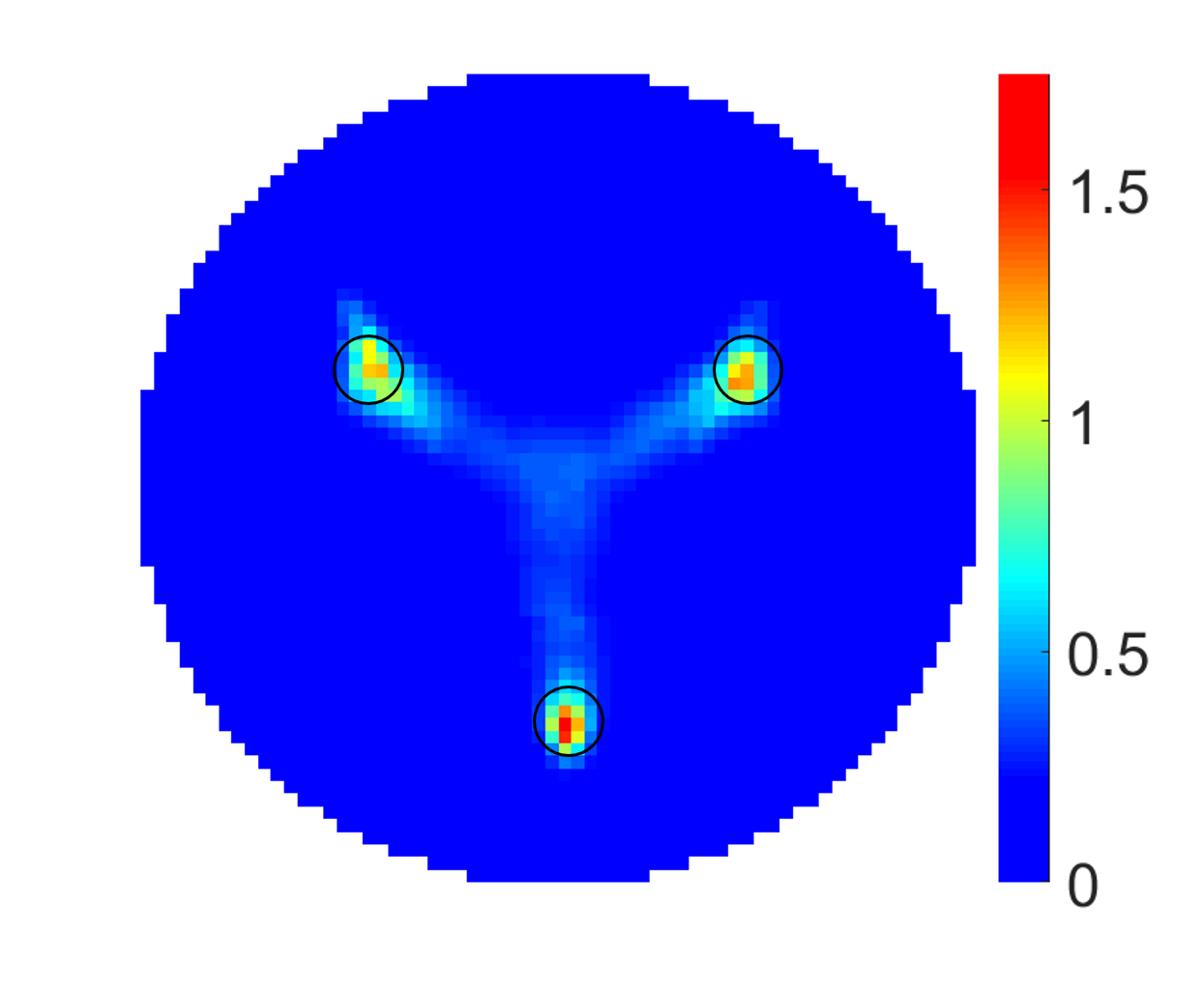} \\
			\hline
		\end{tabular}
	\end{spacing}
	\label{tab:TMSBLPhantom2}
\end{table*}

\begin{table*}[htb]
	\caption{Image reconstruction results based on experimental data (17-mm cylinders, 120-mm sensing region in diameter)}
	\begin{spacing}{1.5}
		\centering
		\begin{tabular}{|ccc|ccc|} 
			\hline 
			\multicolumn{3}{|c|}{Dual Frequency Imaging with SMV} & \multicolumn{3}{|c|}{mfEMT Imaging with MMV}  \\
			\hline
			$f_2 = 1.5625\ MHz$&$f_3 = 3.125\ MHz$&$f_4 = 6.25\ MHz$&$f_2 =1.5625\ MHz$&$f_3 = 3.125\ MHz$&$f_4 = 6.25\ MHz$ \\
			\includegraphics[width=1 in]{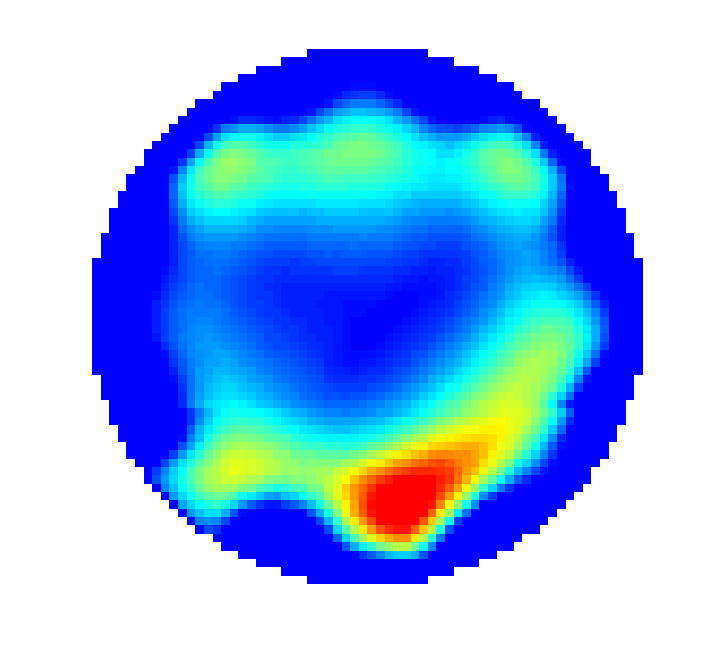} & 
			\includegraphics[width=1 in]{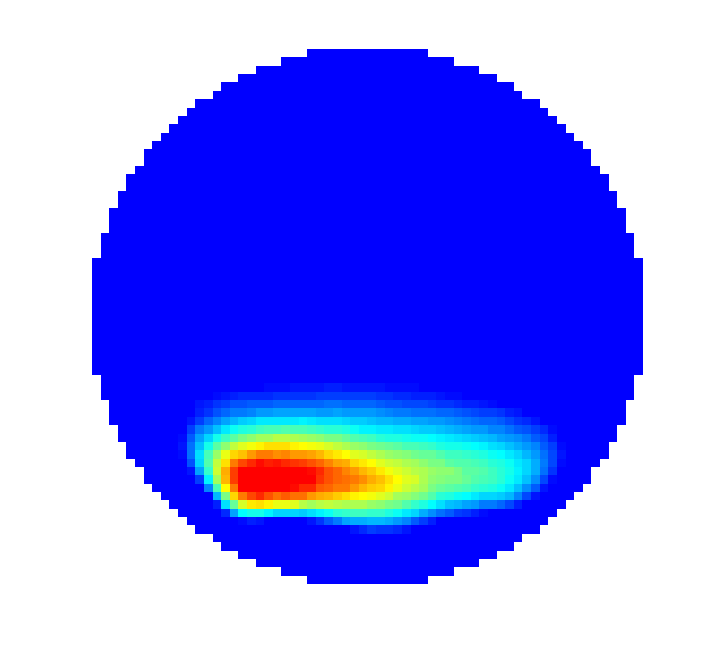} & 
			\includegraphics[width=1 in]{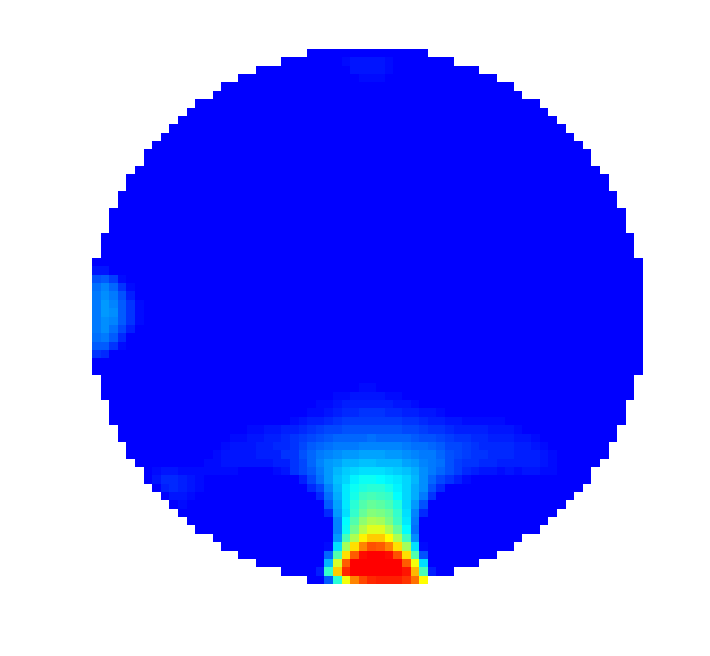} & 
			\includegraphics[width=1 in]{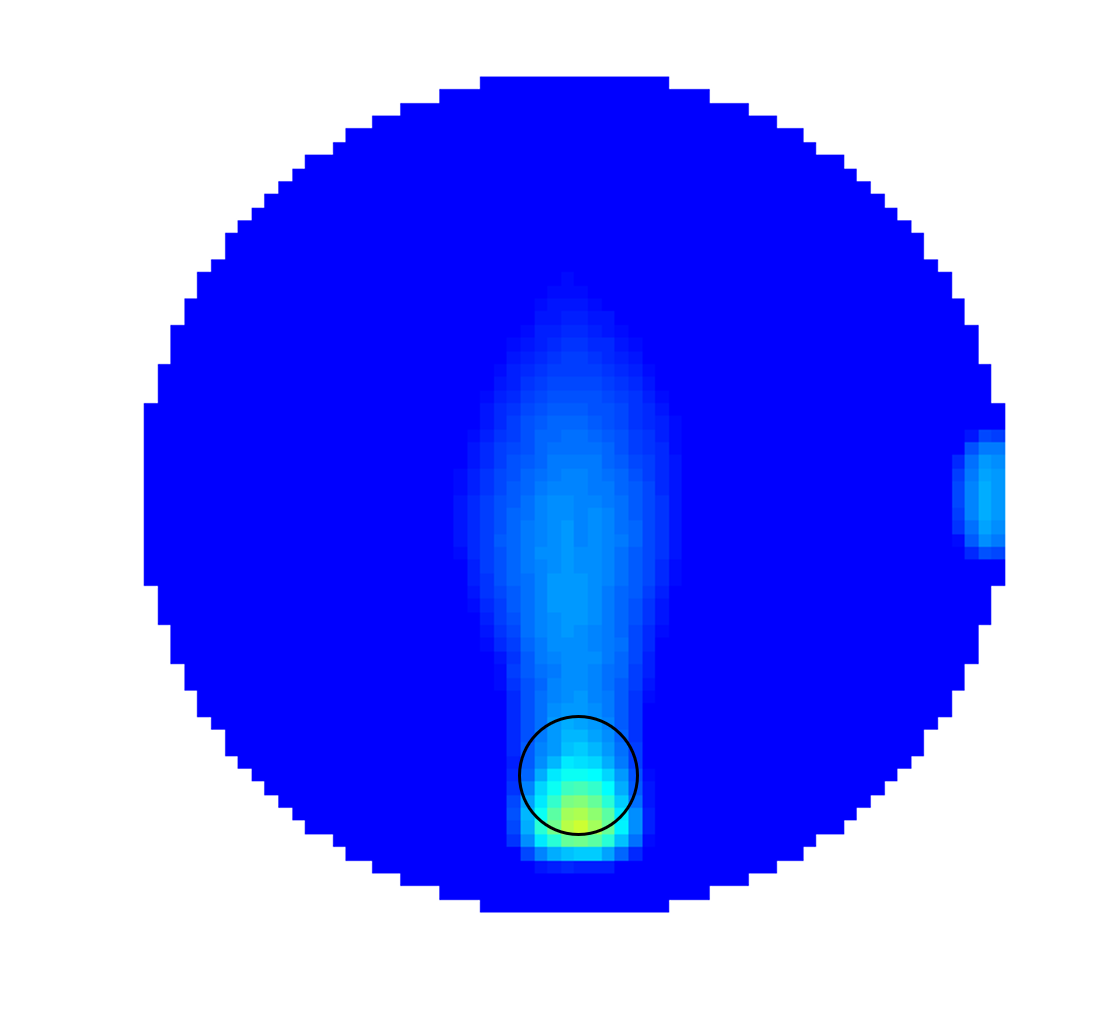} & 
			\includegraphics[width=1 in]{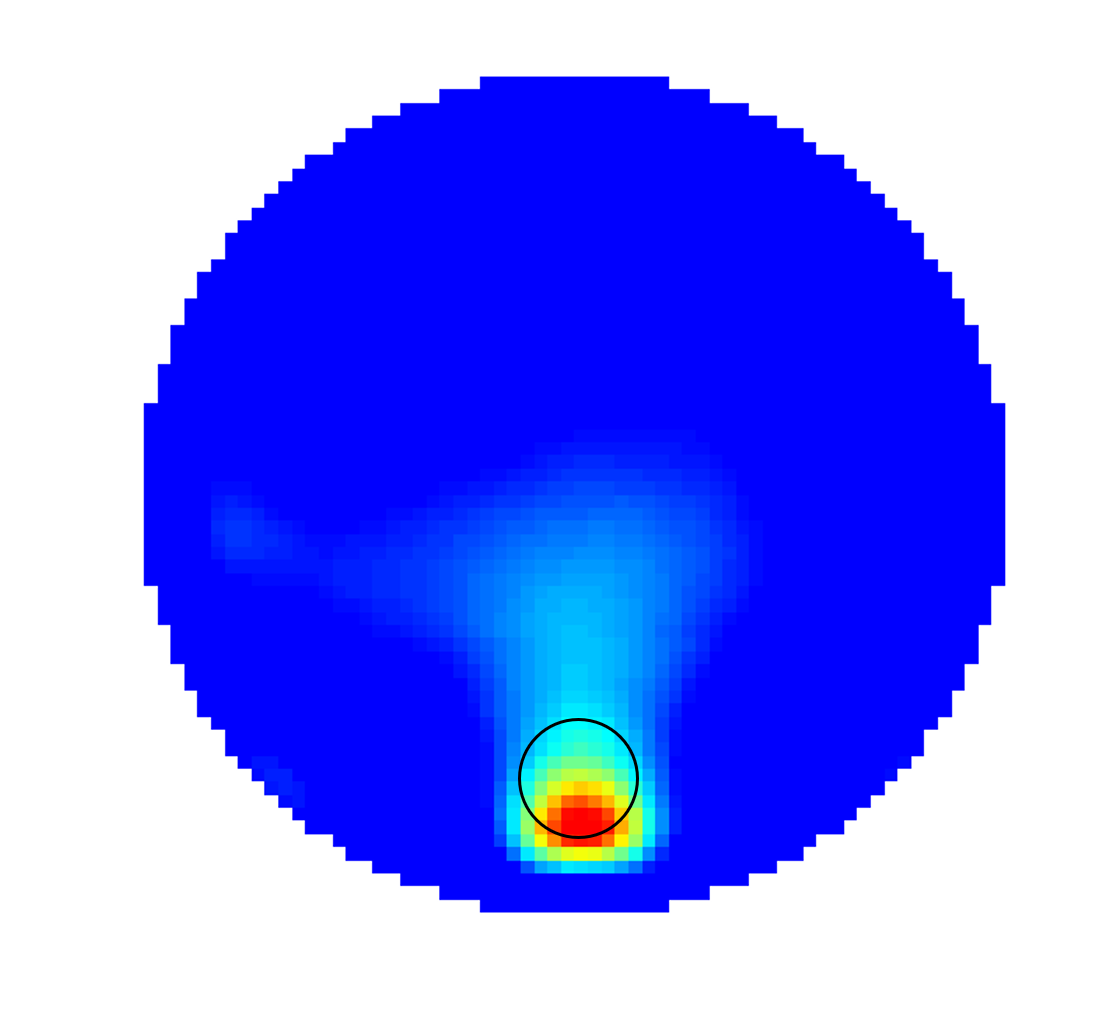} & 
			\includegraphics[width=1.1 in]{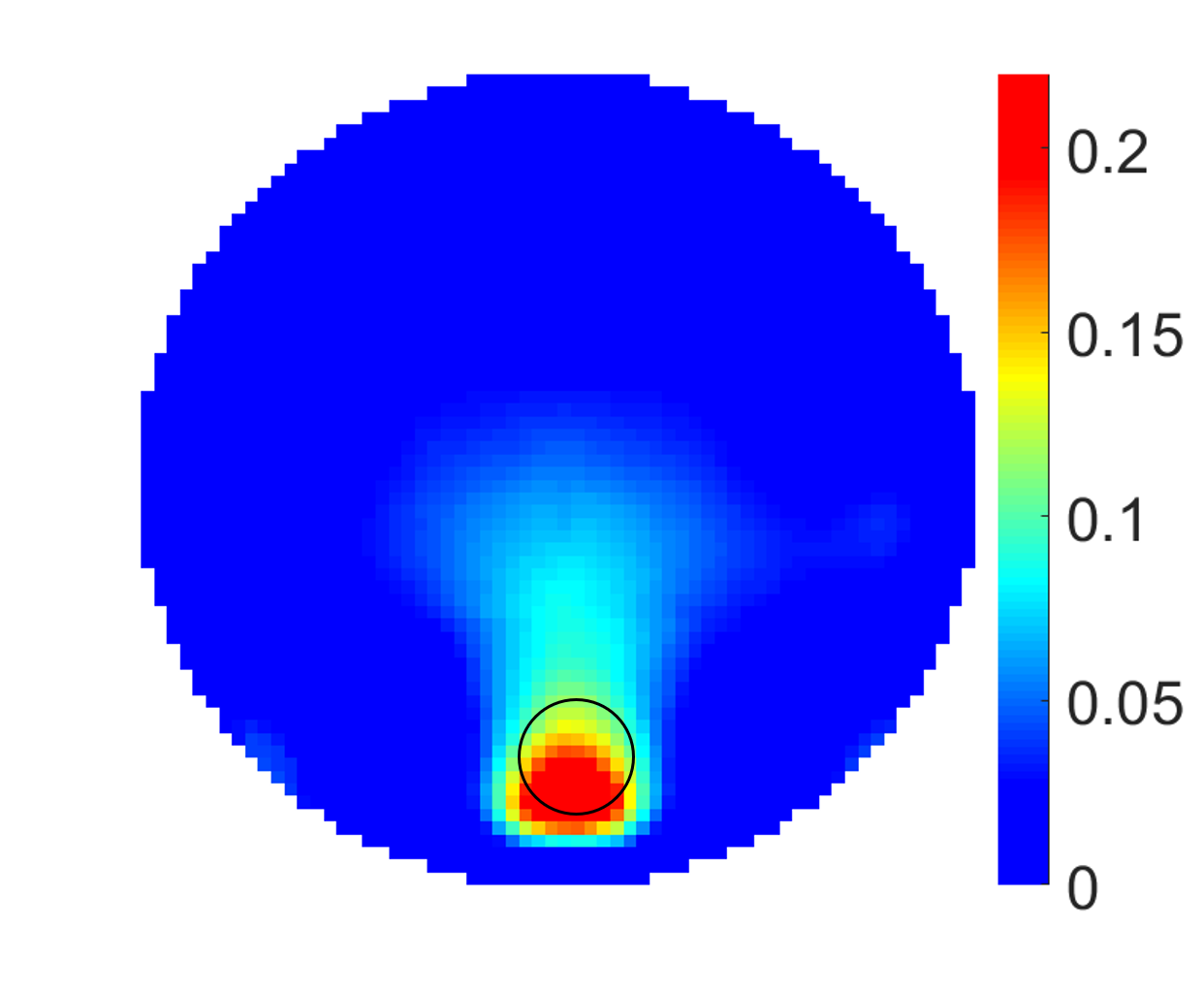} \\
			\includegraphics[width=1 in]{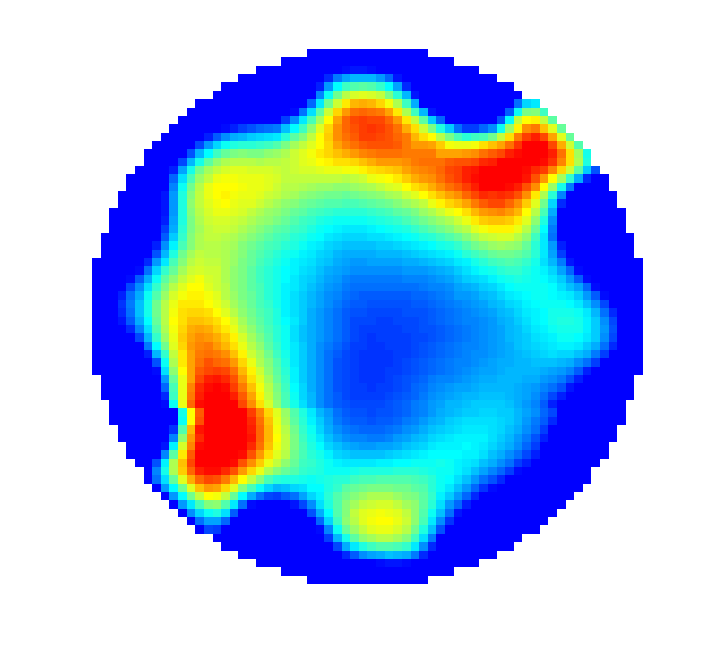} & 
			\includegraphics[width=1 in]{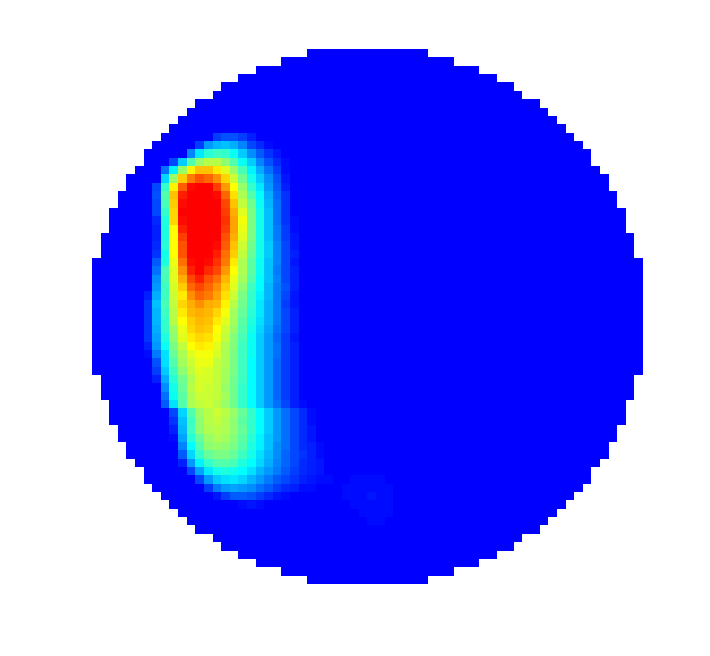} & 
			\includegraphics[width=1 in]{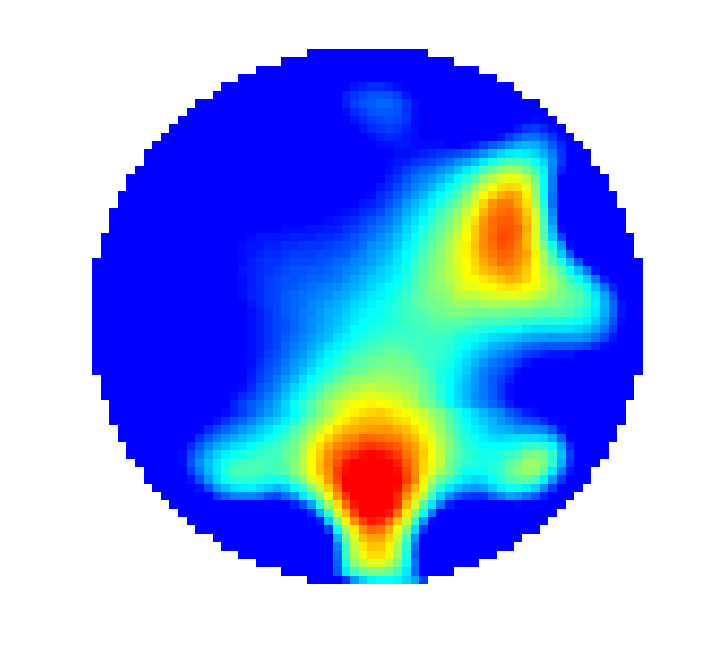} & 
			\includegraphics[width=1 in]{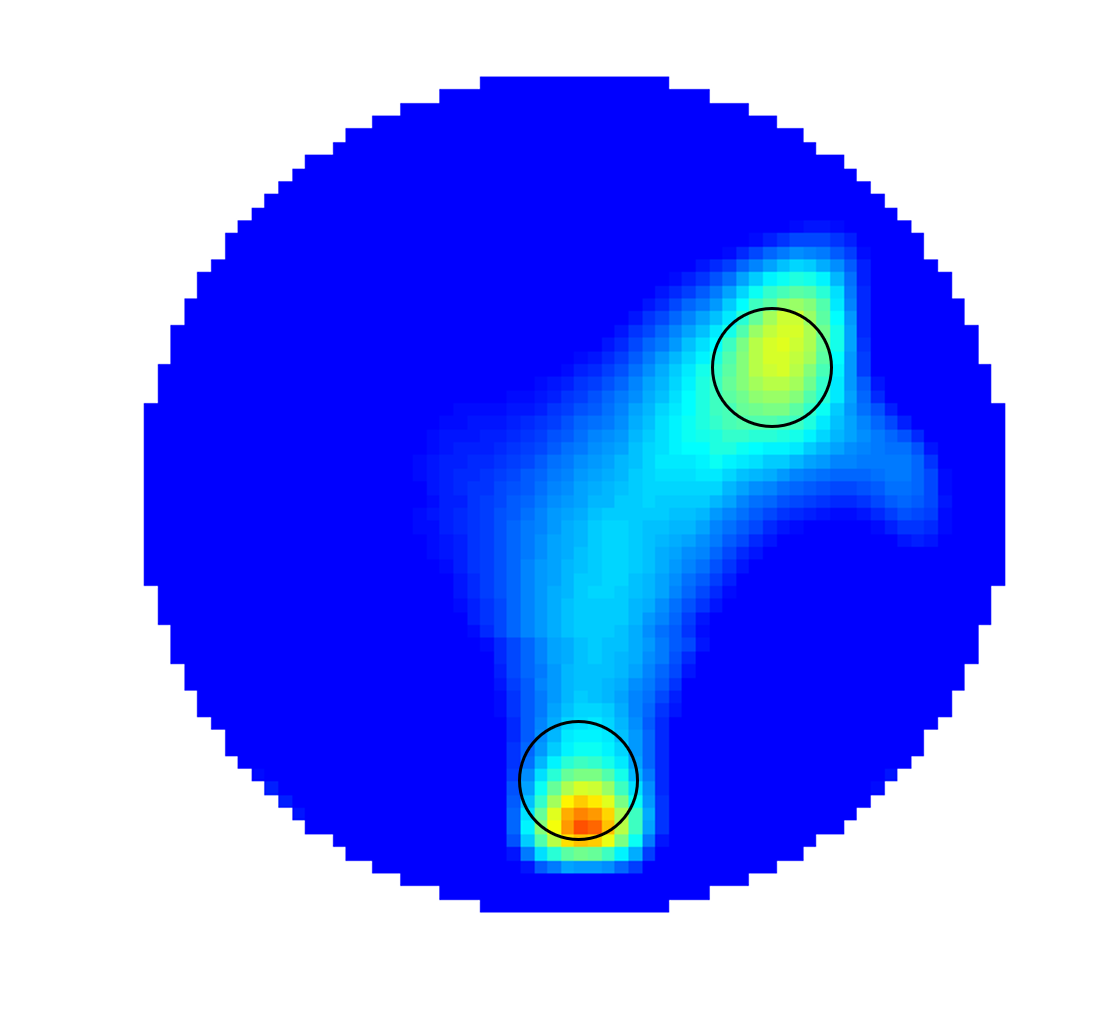} & 
			\includegraphics[width=1 in]{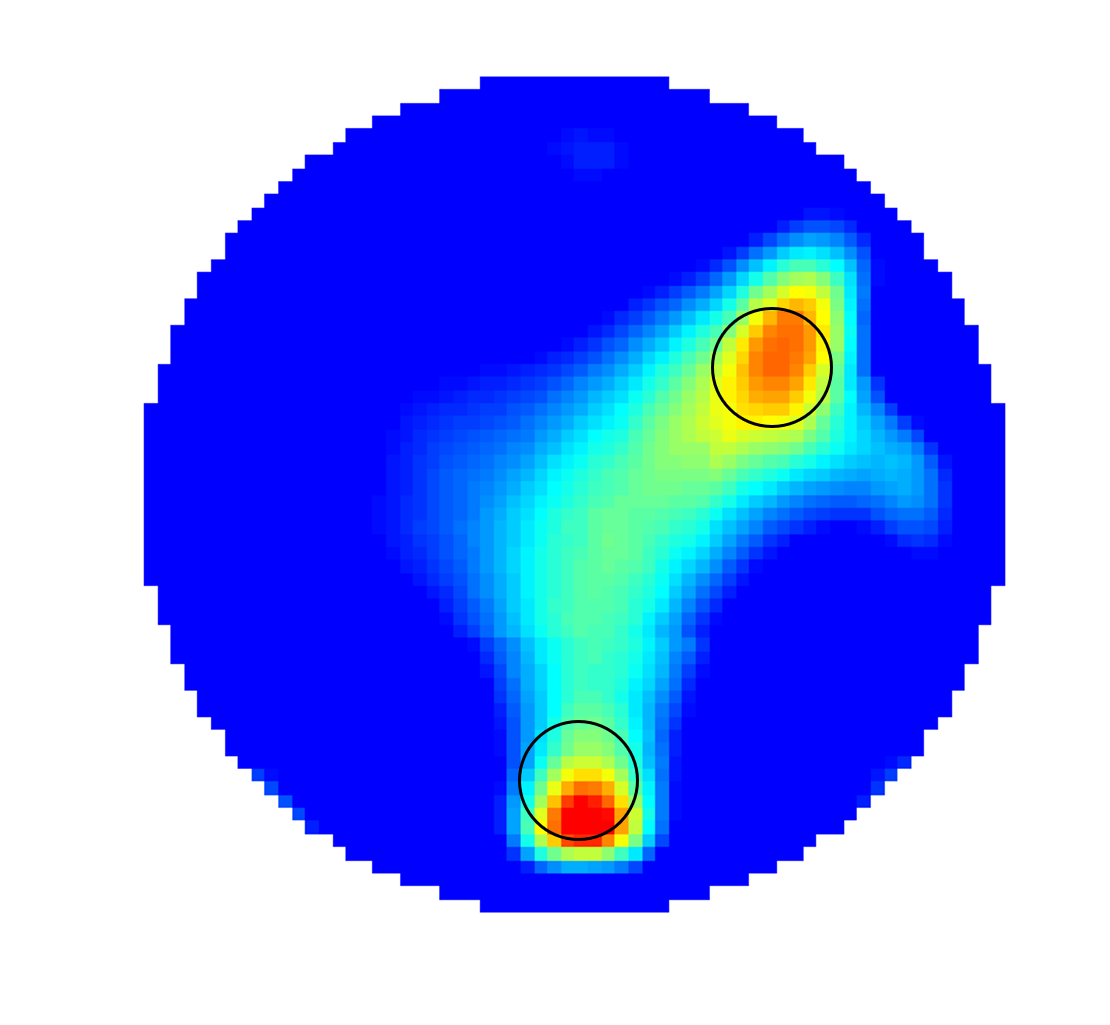} & 
			\includegraphics[width=1.1 in]{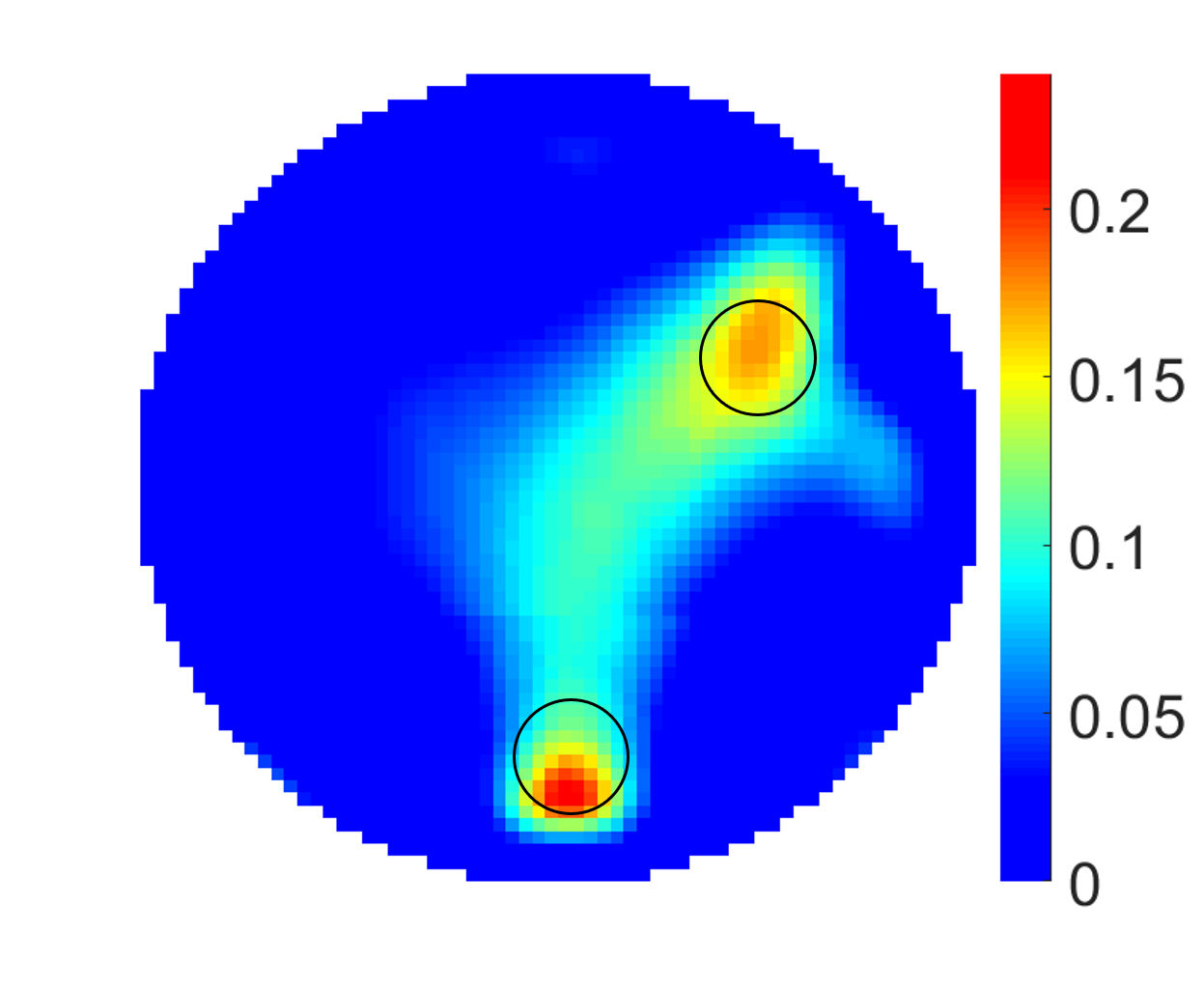} \\
			\includegraphics[width=1 in]{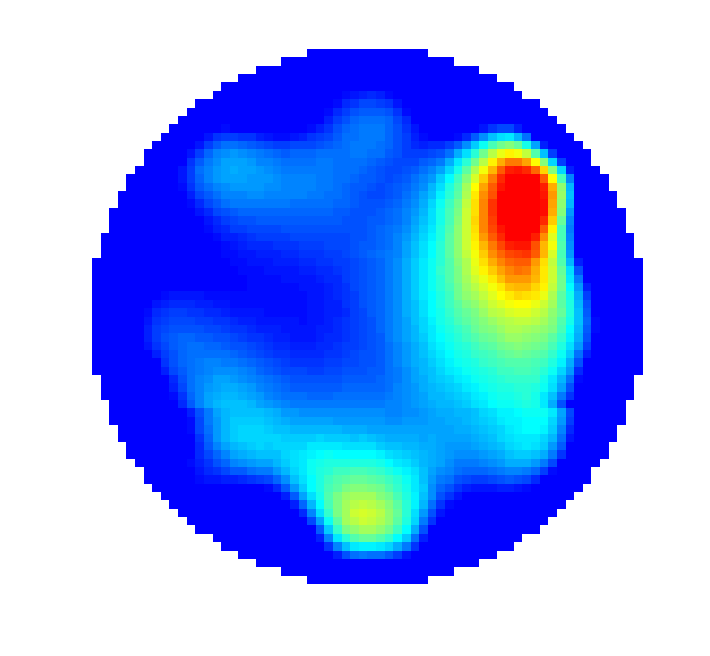} & 
			\includegraphics[width=1 in]{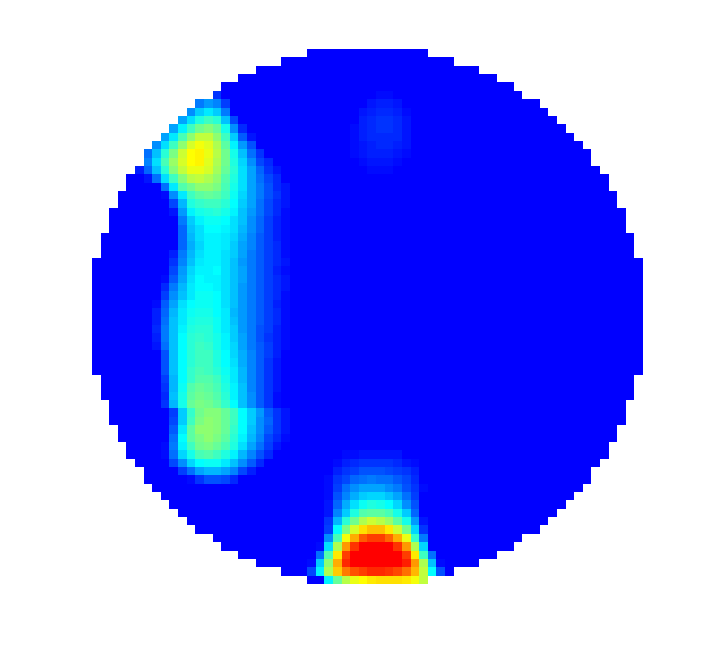} & 
			\includegraphics[width=1 in]{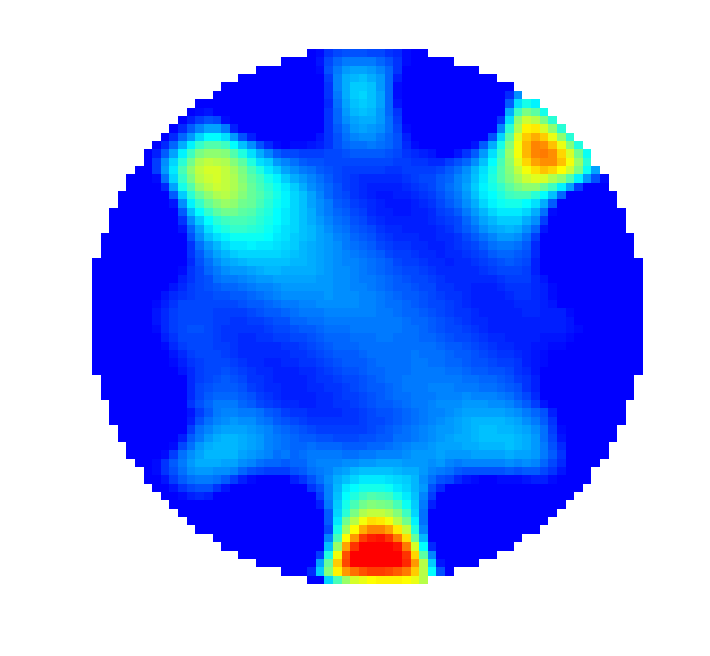} & 
			\includegraphics[width=1 in]{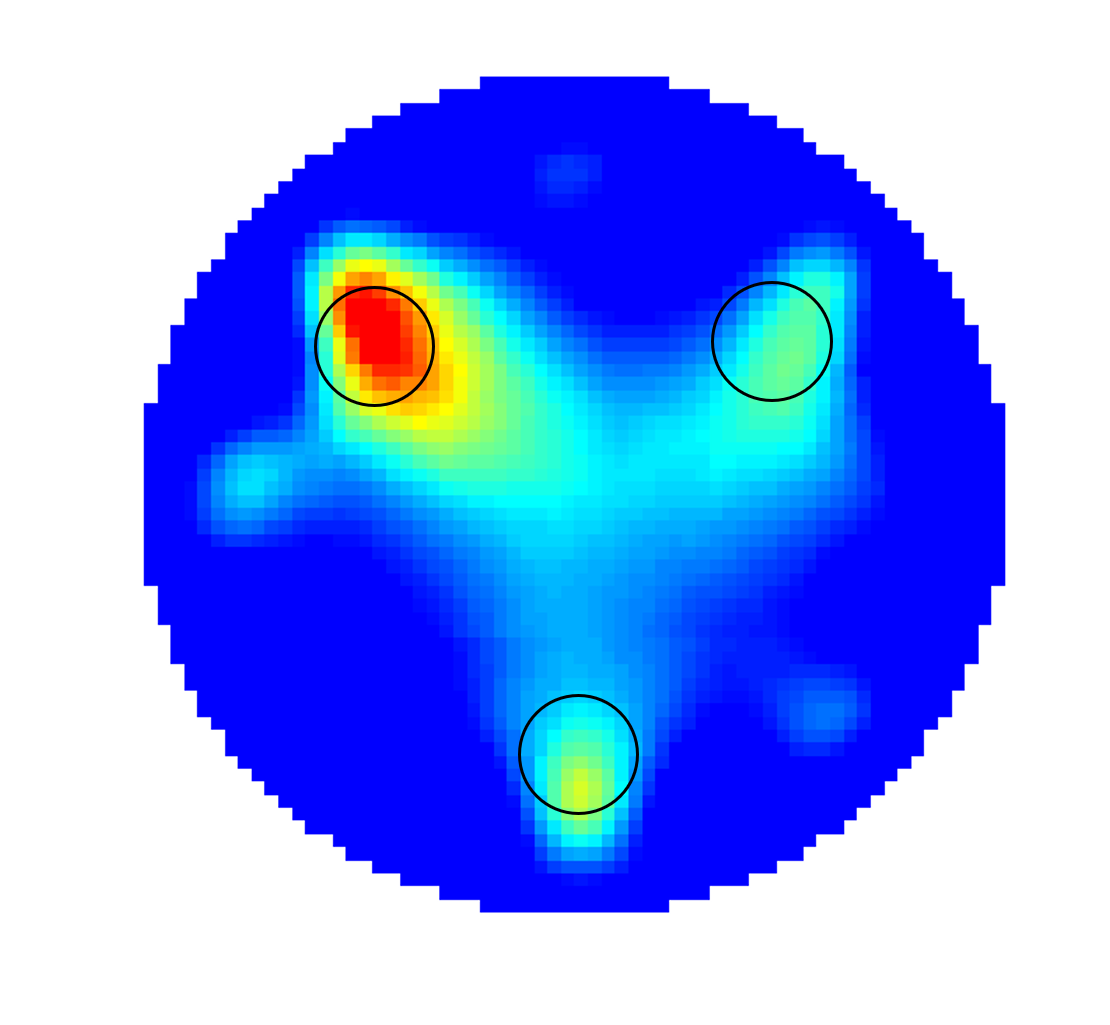} & 
			\includegraphics[width=1 in]{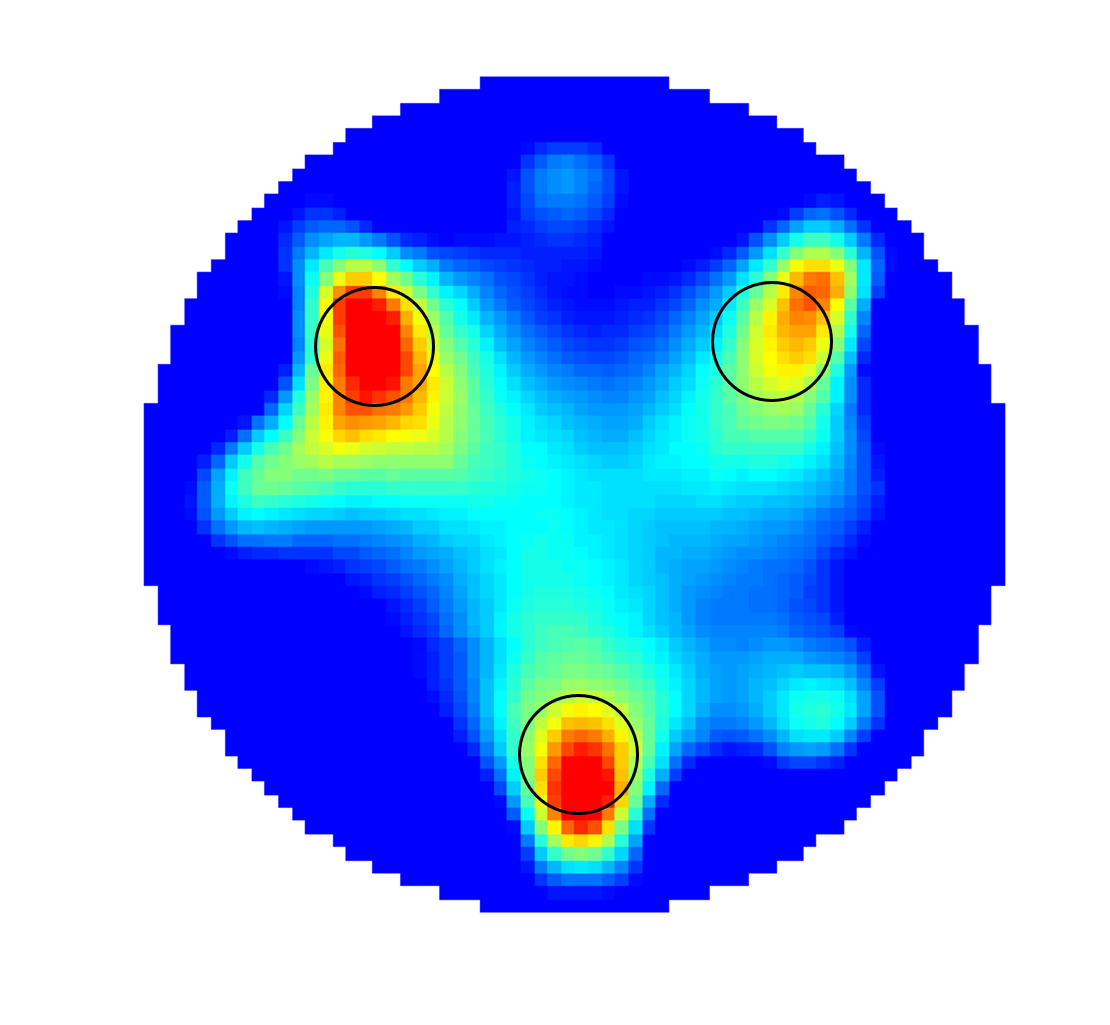} & 
			\includegraphics[width=1.1 in]{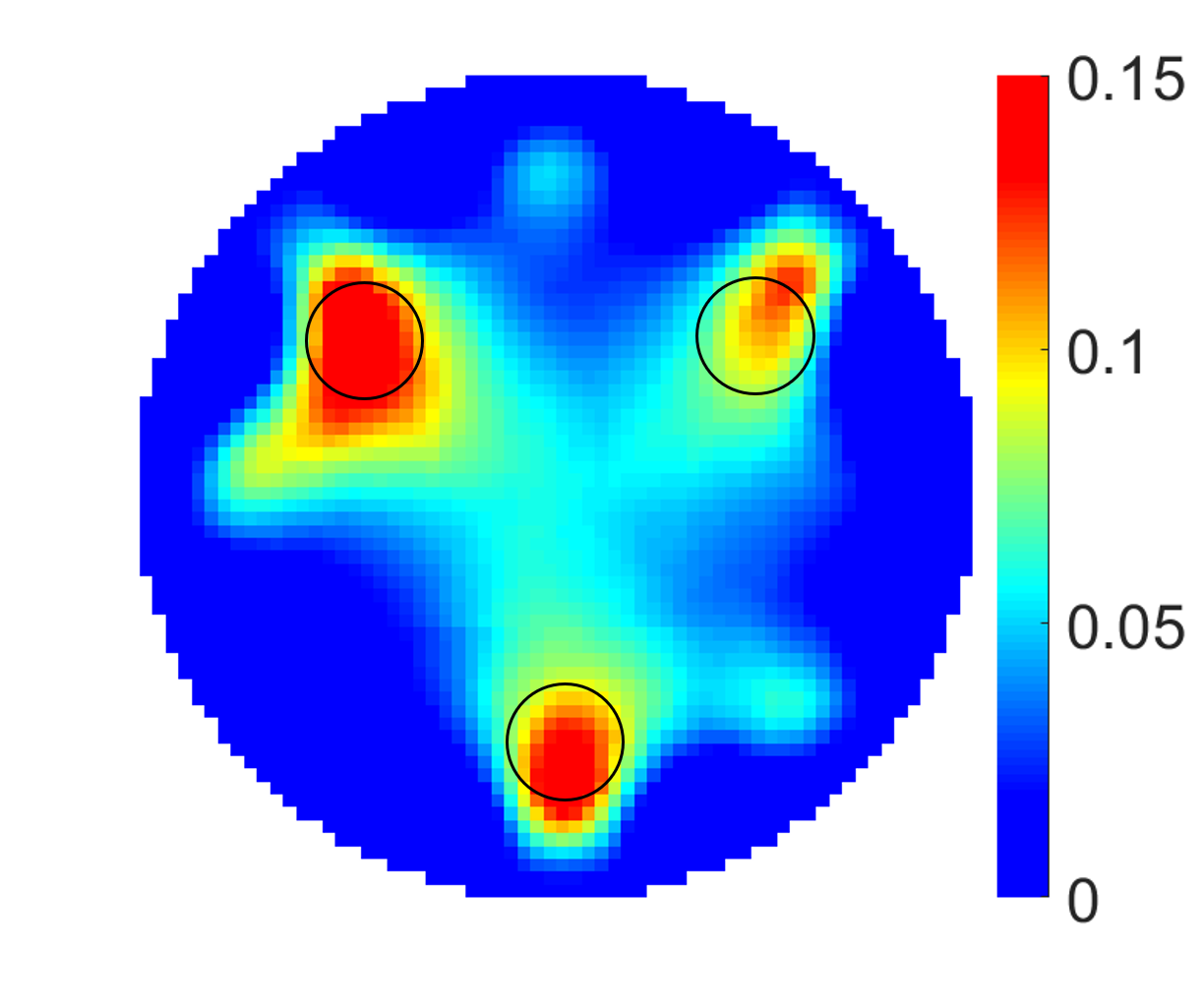} \\
			\hline
		\end{tabular}
	\end{spacing}
	\label{tab:TMSBLPhantom3}
\end{table*}

\section{Conclusion}
\label{sec:conclusion}

This paper presents a novel image reconstruction algorithm for mfEMT by introducing MMV model and sparse Bayesian learning. The proposed method exploits the frequency-related information between each measurement. We introduce the sparse  Bayesian learning method to solve the corresponding MMV problem, which is  especially effective for challenging small scale conductivity distributions. Simulation and experiment have been conducted to verify the performance of the proposed method. Results show that by taking advantage of multiple measurements, it is more robust to noisy data. This feature is crucial  for electromagnetic tomography image reconstruction because the problem is severely ill-posed.


\begin{thebibliography}{10}
	\providecommand{\url}[1]{#1}
	\csname url@samestyle\endcsname
	\providecommand{\newblock}{\relax}
	\providecommand{\bibinfo}[2]{#2}
	\providecommand{\BIBentrySTDinterwordspacing}{\spaceskip=0pt\relax}
	\providecommand{\BIBentryALTinterwordstretchfactor}{4}
	\providecommand{\BIBentryALTinterwordspacing}{\spaceskip=\fontdimen2\font plus
		\BIBentryALTinterwordstretchfactor\fontdimen3\font minus
		\fontdimen4\font\relax}
	\providecommand{\BIBforeignlanguage}[2]{{%
			\expandafter\ifx\csname l@#1\endcsname\relax
			\typeout{** WARNING: IEEEtran.bst: No hyphenation pattern has been}%
			\typeout{** loaded for the language `#1'. Using the pattern for}%
			\typeout{** the default language instead.}%
			\else
			\language=\csname l@#1\endcsname
			\fi
			#2}}
	\providecommand{\BIBdecl}{\relax}
	\BIBdecl
	
	\bibitem{o2015non}
	M.~D. O'Toole, L.~A. Marsh, J.~L. Davidson, Y.~M. Tan, D.~W. Armitage, and
	A.~J. Peyton, ``Non-contact multi-frequency magnetic induction spectroscopy
	system for industrial-scale bio-impedance measurement,'' \emph{Measurement
		Science and Technology}, vol.~26, no.~3, p. 035102, 2015.
	
	\bibitem{pliquett2010bioimpedance}
	U.~Pliquett, ``Bioimpedance: a review for food processing,'' \emph{Food
		engineering reviews}, vol.~2, no.~2, pp. 74--94, 2010.
	
	\bibitem{soltani2011evaluating}
	M.~Soltani, R.~Alimardani, and M.~Omid, ``Evaluating banana ripening status
	from measuring dielectric properties,'' \emph{Journal of Food Engineering},
	vol. 105, no.~4, pp. 625--631, 2011.
	
	\bibitem{justice2011process}
	C.~Justice, A.~Brix, D.~Freimark, M.~Kraume, P.~Pfromm, B.~Eichenmueller, and
	P.~Czermak, ``Process control in cell culture technology using dielectric
	spectroscopy,'' \emph{Biotechnology advances}, vol.~29, no.~4, pp. 391--401,
	2011.
	
	\bibitem{xiao2018multi}
	Z.~Xiao, C.~Tan, and F.~Dong, ``Multi-frequency difference method for
	intracranial hemorrhage detection by magnetic induction tomography,''
	\emph{Physiological measurement}, vol.~39, no.~5, p. 055006, 2018.
	
	\bibitem{yang2017multi}
	Y.~Yang and J.~Jia, ``A multi-frequency electrical impedance tomography system
	for real-time 2d and 3d imaging,'' \emph{Review of Scientific Instruments},
	vol.~88, no.~8, p. 085110, 2017.
	
	\bibitem{wang2017magnetic}
	J.-Y. Wang, T.~Healey, A.~Barker, B.~Brown, C.~Monk, and D.~Anumba, ``Magnetic
	induction spectroscopy (mis)—probe design for cervical tissue
	measurements,'' \emph{Physiological measurement}, vol.~38, no.~5, p. 729,
	2017.
	
	\bibitem{jiang2018capacitively}
	Y.~Jiang and M.~Soleimani, ``Capacitively coupled resistivity imaging for
	biomaterial and biomedical applications,'' \emph{IEEE Access}, vol.~6, pp.
	27\,069--27\,079, 2018.
	
	\bibitem{yang2011multiple}
	J.~Yang, A.~Bouzerdoum, F.~H.~C. Tivive, and M.~G. Amin, ``Multiple-measurement
	vector model and its application to through-the-wall radar imaging,'' in
	\emph{2011 IEEE International Conference on Acoustics, Speech and Signal
		Processing (ICASSP)}.\hskip 1em plus 0.5em minus 0.4em\relax IEEE, 2011, pp.
	2672--2675.
	
	\bibitem{ziniel2012efficient}
	J.~Ziniel and P.~Schniter, ``Efficient high-dimensional inference in the
	multiple measurement vector problem,'' \emph{IEEE Transactions on Signal
		Processing}, vol.~61, no.~2, pp. 340--354, 2012.
	
	\bibitem{yang2012off}
	Z.~Yang, L.~Xie, and C.~Zhang, ``Off-grid direction of arrival estimation using
	sparse bayesian inference,'' \emph{IEEE Transactions on Signal Processing},
	vol.~61, no.~1, pp. 38--43, 2012.
	
	\bibitem{majumdar2011accelerating}
	A.~Majumdar and R.~K. Ward, ``Accelerating multi-echo t2 weighted mr imaging:
	analysis prior group-sparse optimization,'' \emph{journal of Magnetic
		Resonance}, vol. 210, no.~1, pp. 90--97, 2011.
	
	\bibitem{xiang2019design}
	J.~Xiang, Y.~Dong, M.~Zhang, and Y.~Li, ``Design of a magnetic induction
	tomography system by gradiometer coils for conductive fluid imaging,''
	\emph{IEEE Access}, vol.~7, pp. 56\,733--56\,744, 2019.
	
	\bibitem{REDPITAYA}
	\BIBentryALTinterwordspacing
	Redpitaya.com. (2019) Red pitaya. [Online]. Available:
	\url{https://www.redpitaya.com/}
	\BIBentrySTDinterwordspacing
	
	\bibitem{rosell2001sensitivity}
	J.~Rosell, R.~Casanas, and H.~Scharfetter, ``Sensitivity maps and system
	requirements for magnetic induction tomography using a planar gradiometer,''
	\emph{Physiological measurement}, vol.~22, no.~1, p. 121, 2001.
	
	\bibitem{liu2018image}
	S.~Liu, J.~Jia, Y.~D. Zhang, and Y.~Yang, ``Image reconstruction in electrical
	impedance tomography based on structure-aware sparse bayesian learning,''
	\emph{IEEE transactions on medical imaging}, vol.~37, no.~9, pp. 2090--2102,
	2018.
	
	\bibitem{zhang2011sparse}
	Z.~Zhang and B.~D. Rao, ``Sparse signal recovery with temporally correlated
	source vectors using sparse bayesian learning,'' \emph{IEEE Journal of
		Selected Topics in Signal Processing}, vol.~5, no.~5, pp. 912--926, 2011.
	
	\bibitem{wipf2007empirical}
	D.~P. Wipf and B.~D. Rao, ``An empirical bayesian strategy for solving the
	simultaneous sparse approximation problem,'' \emph{IEEE Transactions on
		Signal Processing}, vol.~55, no.~7, pp. 3704--3716, 2007.
	
\end{thebibliography}
\end{document}